\documentclass[twocolumn,a4,showpacs,preprintnumbers,amsmath,amssymb]{revtex4-1}
\usepackage{color}
\usepackage{epsfig}

\usepackage{mathptmx, courier, pifont}
\usepackage[scaled=0.92]{helvet}
\usepackage[T1]{fontenc}
\usepackage{textcomp}

\usepackage[pdftex,colorlinks=true]{hyperref}

 \def\ket{\!>\,} \def\ack{\,|\,}
\begin{document}

\author{\large S. Jehangir$^{1}$}
\author{\large G.H. Bhat$^{2,3,4}$}
\author{\large N. Rather$^1$}
\author{\large J.A. Sheikh$^{4}$}
\author{\large R. Palit$^5$}
\affiliation{$^1$Department of Physics, Islamic University of Science and Technology,  Jammu and Kashmir, 192 122, India}
\affiliation{$^{2}$ Department of Physics, S.P. College,  Srinagar, Jammu and Kashmir, 190001, India}
\affiliation{$^3$Cluster University Srinagar, Jammu and Kashmir,
  Srinagar, Goji Bagh, 190008, India}
\affiliation{$^4$Department of Physics, University of Kashmir,
  Hazratbal, Srinagar, 190006, India}
  \affiliation{$^5$Department of Nuclear and Atomic Physics, 
Tata Institute of Fundamental Research, Mumbai - 400005, India}

\title{ Systematic study of  near yrast band structures in odd-mass $^{125-137}$Pr and $^{127-139}$Pm isotopes}

\begin{abstract}
In the present work, the basis space in the triaxial projected shell model
approach is expanded to include three and five quasiparticle
configurations for odd-proton systems. This extension allows to 
investigate the high-spin band structures observed in odd-proton systems up to
and including the second band crossing region, and as a first major application of
this development, the high-spin properties are investigated for
odd-mass $^{125-137}$Pr and $^{127-139}$Pm isotopes. It is
shown that band crossings in the studied isotopes have mixed structures with first
crossing dominated by one-proton coupled to two-neutron configuration for the
lighter isotopes which then changes to three-proton configuration with increasing
neutron number. Further, $\gamma$-bands based on quasiparticle states are also 
delineated in the present work, and it is predicted that these band structures built on
three-quasiparticle configurations become favoured in energy for heavier systems in the high-spin region.
\end{abstract}

\date{\today}


\maketitle
\section{\label{sec:Intro}Introduction}
 Nuclei in the mass $\sim$ 130 region are known to exhibit a rich variety of
shapes and structures.  In this region, interesting phenomena of
shape co-existence \cite{coe1, coe2}, strongly deformed \cite{hd2} to
superdeformed \cite{sd1, sd2} shapes, chiral doublet bands \cite{cdb1, cdb2} and $\gamma$-
bands built on quasiparticle states \cite{xee,xe,xe1} have
been observed. This is the heaviest mass region with valence neutrons and
protons occupying the same intruder orbital, $1h_{11/2}$.  For the
neutron-deficient isotopic chains in this mass region, protons occupy the
low-$\Omega$ orbitals, whereas neutrons occupancy changes from
mid-$\Omega$ to high-$\Omega$ orbitals of $1h_{11/2}$. Due to the
competing shape polarising effects of  low-$\Omega$ and high-$\Omega$ orbitals, the
neutron-deficient nuclei in this region are expected to have, in general, triaxial
shapes \cite{AG96,RW02}. 

 \begin{table*}
\caption{Axial and triaxial quadrupole deformation parameters
$\epsilon$ and $\epsilon'$  employed in the TPSM calculation.  }
\begin{tabular}{|ccccccccccccccc|}
\hline                   & $^{125}$Pr &  $^{127}$Pr    &  $^{129}$Pr  &$^{131}$Pr  & $^{133}$Pr& $^{135}$Pr  & $^{137}$Pr  &$^{127}$Pm & $^{129}$Pm  &$^{131}$Pm  &$^{133}$Pm &  $^{135}$Pm & $^{137}$Pm & $^{139}$Pm \\
\hline   $\epsilon$      &0.300      &  0.283       &  0.267      & 0.234     & 0.194    &  0.150    &  0.150     &    0.300  &   0.300    &   0.300   &   0.292  &    0.230  &  0.200   &    0.190  \\
\hline $\epsilon'$       &0.100      &  0.100       &  0.100      & 0.101     & 0.090    &  0.080    &  0.080     &    0.110  &   0.110    &   0.110   &   0.120  &    0.110  &  0.100   &    0.090  \\
\hline $\gamma$          & 18.4      &  19.5        & 20.5        & 23.3      & 24.8     & 28.1      & 28.1       &   20.1    &   20.1     & 20.1      & 22.3     &    25.6   & 26.5     &  25.3     \\
\hline
\end{tabular}\label{tab1}
\end{table*}

\begin{figure*}[htb]
\vspace{0cm}
\includegraphics[totalheight=14cm]{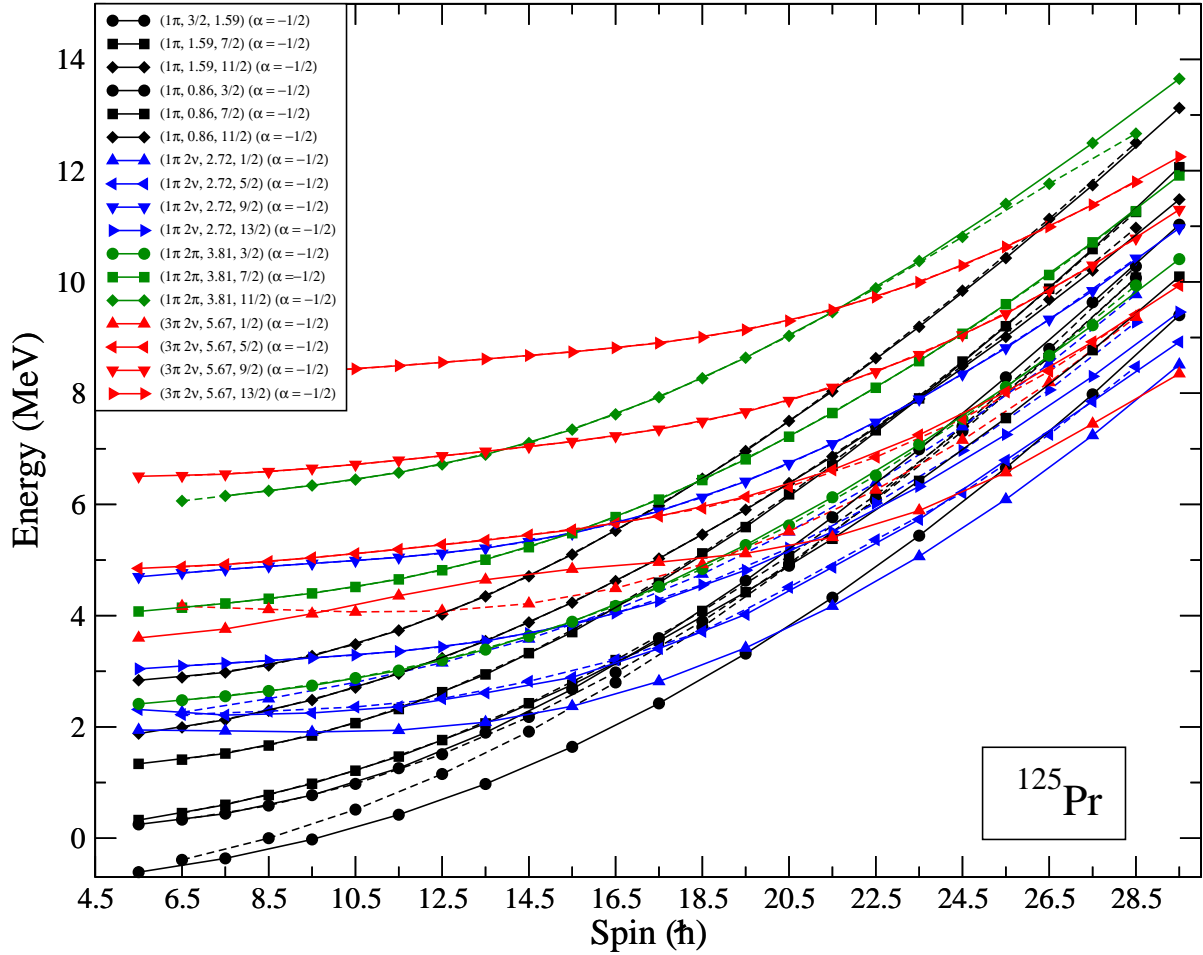} \caption{(Color
online)  Projected energies are shown before diagonalization of the
shell model Hamiltonian for $^{125}$Pr. The bands are labelled by
three quantities : group structure, energy and K-quantum number of the quasiparticle
state. For instance, $(1\pi,1.54,3/2)$ designates one-quasiproton
state having intrinsic energy of 1.54 MeV and K$=3/2$. The two
signature bands for low-K states are depicted separately as the energy
splitting between the two branches is large and the plots become   
quite clumsy when plotted as a single curve. In the legend of the
figure bands are designated for $\alpha=-1/2$ states and for the
$\alpha=+1/2$ states same symbols are used except that the
corresponding curves are dashed lines.  } \label{bd1}
\end{figure*}
\begin{figure*}[htb]
\vspace{0cm}
\includegraphics[totalheight=15cm]{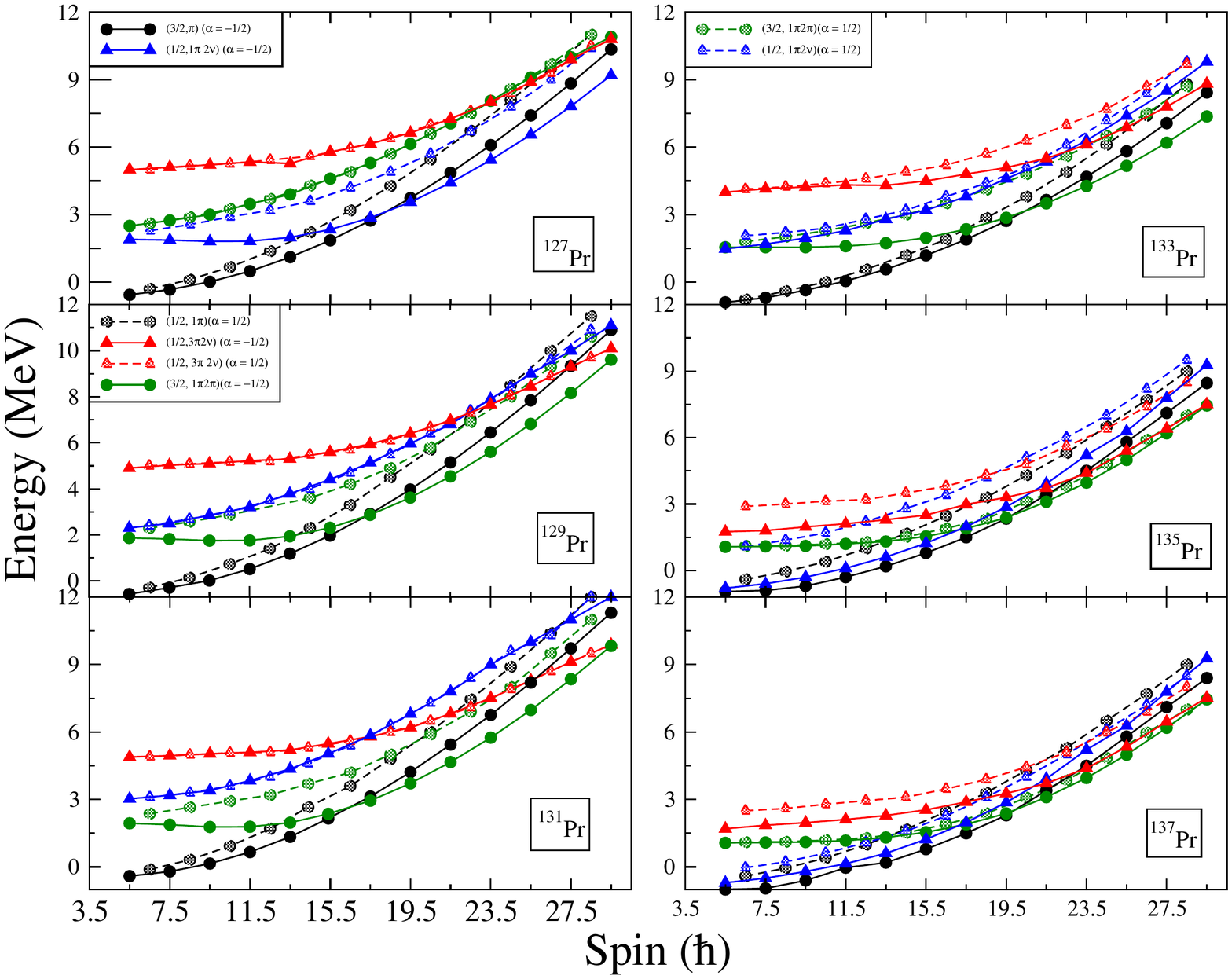} \caption{(Color
online)  Band diagrams for odd-proton $^{127-137}$Pr  isotopes. The
bands are labelled as in Fig.~\ref{bd1} with solid lines representing
$\alpha = -1/2$ and the dashed curves designating $\alpha =
+1/2$. Only the potions of diagrams that encompasses the band crossing are displayed.} \label{bd3}
\end{figure*}
The interplay between proton and neutron configurations also plays an
important role in the elucidation of the high-spin band structures observed
in this mass region. Band structures have been observed up
to quite high-spin and band crossing features have attracted a considerable
attention \cite{ml06,sz00,qx08}.
In particular, the nature of the band crossings in odd-proton Pr- and
Pm-isotopes  has been extensively studied in recent years
\cite{fs21,an02,es11,ad12,bcr}.
It has been shown that standard cranked shell
model (CSM) approach with fixed pairing and deformation fields can
describe the band crossing features reasonably well for heavier Pr- and
Pm-isotopes. However, for lighter isotopes of $^{127}$Pr and
$^{131}$Pm, the gain in alignment is substantially underpredicted using
this approach \cite{cm98}.
The band crossings in these nuclei have also been investigated using the extended version of
total Routhian surface (TRS) approach \cite{cm98} in which pairing and 
deformation fields are determined self-consistently. The observed band
crossing features have been reproduced in this more realistic approach,
and it has been demonstrated that nature of the first band crossing is quite
different from that predicted using the standard CSM approach. It has
been shown that for lighter isotopes band crossings for these isotopes
have dominant contribution from neutron
configuration. This is in contradiction to the standard CSM results which
predict proton BC crossing earlier than the neutron AB crossing for these
nuclei.  

Further, band crossing features in odd-proton isotopes have been
investigated using the projected shell model (PSM) approach. In this
model, basis states are constructed from the solutions of the 
Nilsson potential with axial symmetry \cite{ysm}.  In the study of
odd-proton nuclei, the basis space in PSM is comprised of one-proton and
one-proton coupled to two-neutron configurations. It has been shown
using this approach that band crossing features of lighter isotopes of Promethium
could be described well. However, for heavier isotopes discrepancies
were observed between the PSM predicted and the experimental data.
The major reason for this discrepancy is due to neglect of the
proton aligning configurations in the basis space of PSM since it is 
evident from the CSM analysis\cite{cm98} that proton contribution becomes
more dominant for heavier Pr- and Pm- isotopes. In order to 
elucidate the band crossing features for these isotopes, it is
imperative to include both neutron- and proton-aligning configurations
in the basis space. 
In the present work, we have generalized
the basis space of the projected shell model for odd-proton systems by
including proton aligning configurations, in addition, to the neutron
states. The generalised basis configuration space has
been implemented in the three-dimensional version of the projected shell model,
what is now referred to as the triaxial projected shell model (TPSM) as
most of the Pr- and Pm- isotopes discussed in the present work are 
predicted to have triaxial shapes. We have also included
five-quasiparticle configurations in the basis space, which
allows to investigate the second band crossing observed in some of
these isotopes.
\begin{figure*}[htb]
\vspace{1cm}
\includegraphics[totalheight=14cm]{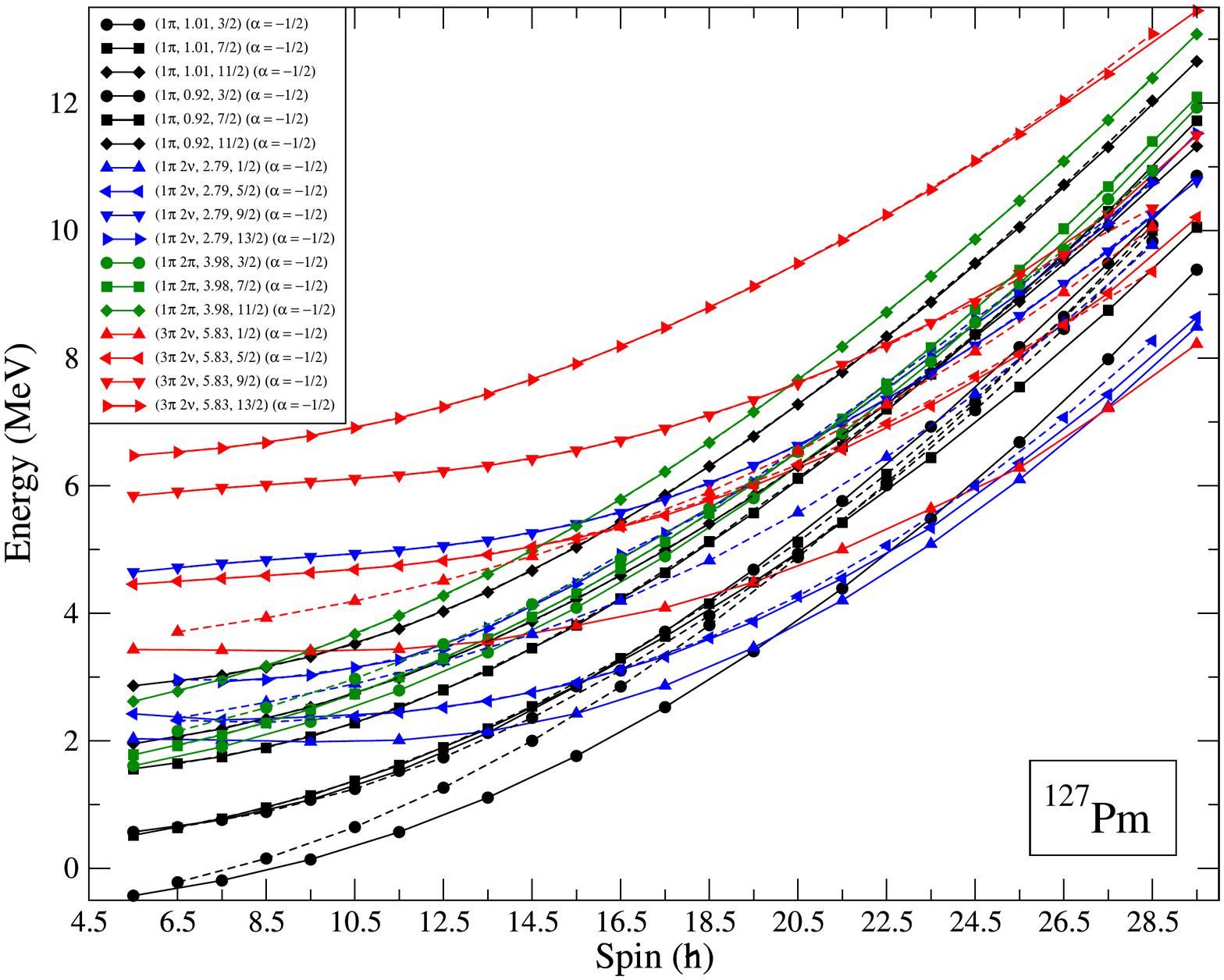} \caption{(Color
online)  Projected energies before diagonalization are shown for
$^{127}$Pm. The labelling of the bands follows the Fig.~\ref{bd1}
description.} \label{bd2}
\end{figure*}
\begin{figure*}[htb]
\vspace{0cm}
\includegraphics[totalheight=15cm]{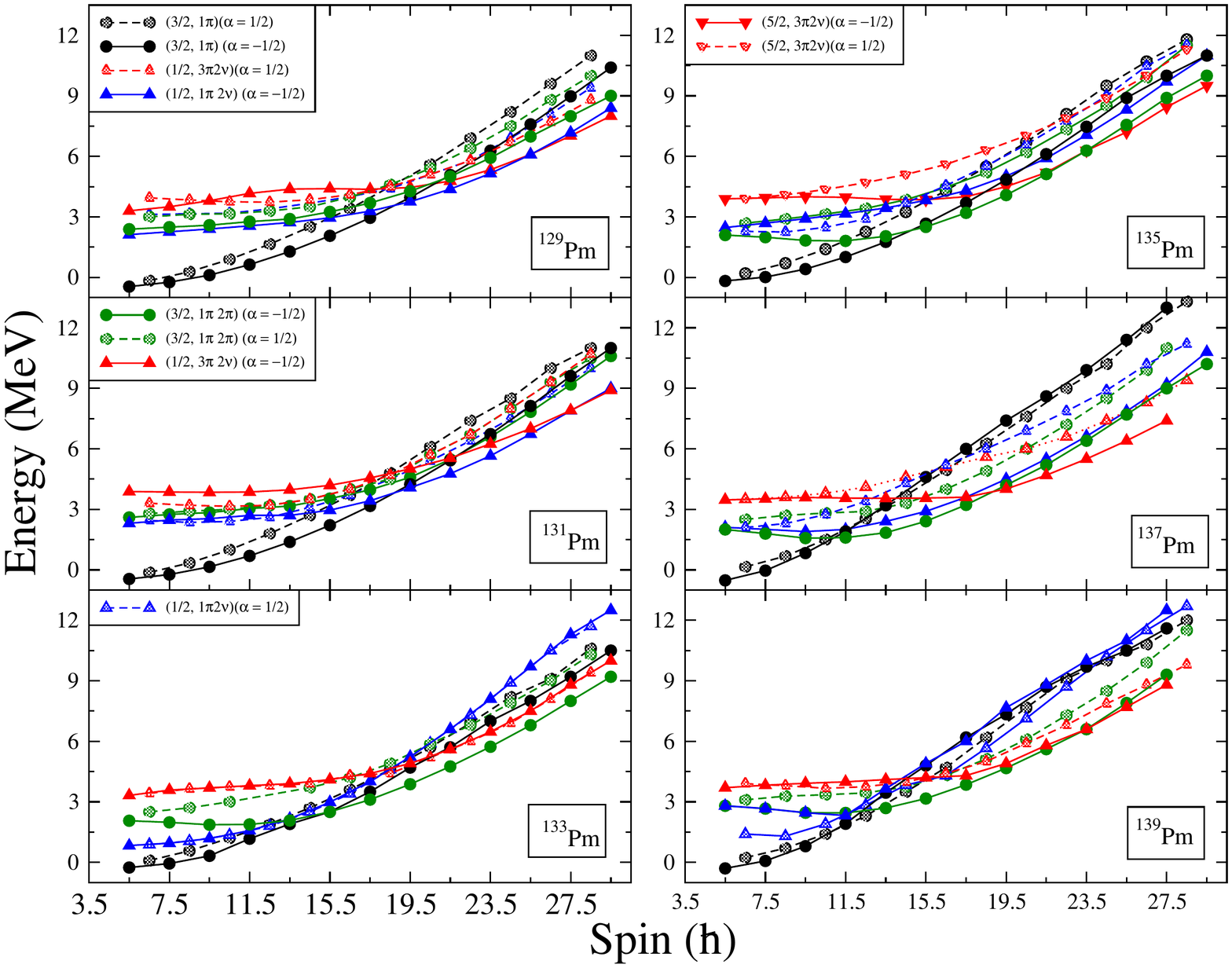} \caption{(Color
online)  Band diagrams for $^{129-139}$Pm  isotopes. The figure is
similar to that of Fig.~\ref{bd3}.} \label{bd4}
\end{figure*}

In recent years, the TPSM approach has turned out to be a useful tool
to investigate the high-spin band structures in deformed and transitional 
nuclei \cite{GH14,an17,sb19,Vs17}.
The model has provided some new insights into nature of the
high-spin band structures in even-even and odd-odd nuclei in  the mass
$\sim$ 130 region. The chiral doublet bands observed in this mass
region have been well described using the TPSM approach \cite{GH14,JG12}.
For a few 
even-even Ce- and Nd- isotopes, it has been
demonstrated that some excited band structures observed are $\gamma$-bands
built on two-quasiparticle states \cite{JG09}. There has been an anomaly
in the g-factor measurements for the band heads of the s-bands
observed in these nuclei \cite{zemel,wyss1,KS87,RW88,RW8,sj17}. The g-factors for the two
observed s-bands
are either positive or negative, implying that the character of both the 
s-bands is either proton or neutron. In mass $\sim$ 130 region, the 
Fermi surfaces of neutrons and protons are close in energy and it is, therefore,
expected that neutrons and protons will align almost simultaneously. From this
perspective, it is expected that two observed s-bands should have 
neutron and proton structures, respectively. The
corresponding g-factors should be positive and negative for the two
s-bands. The observation of both s-bands having positive or negative
g-factors is ruled out using this standard picture. 
\begin{figure*}[htb]
\vspace{1cm}
\includegraphics[totalheight=17cm]{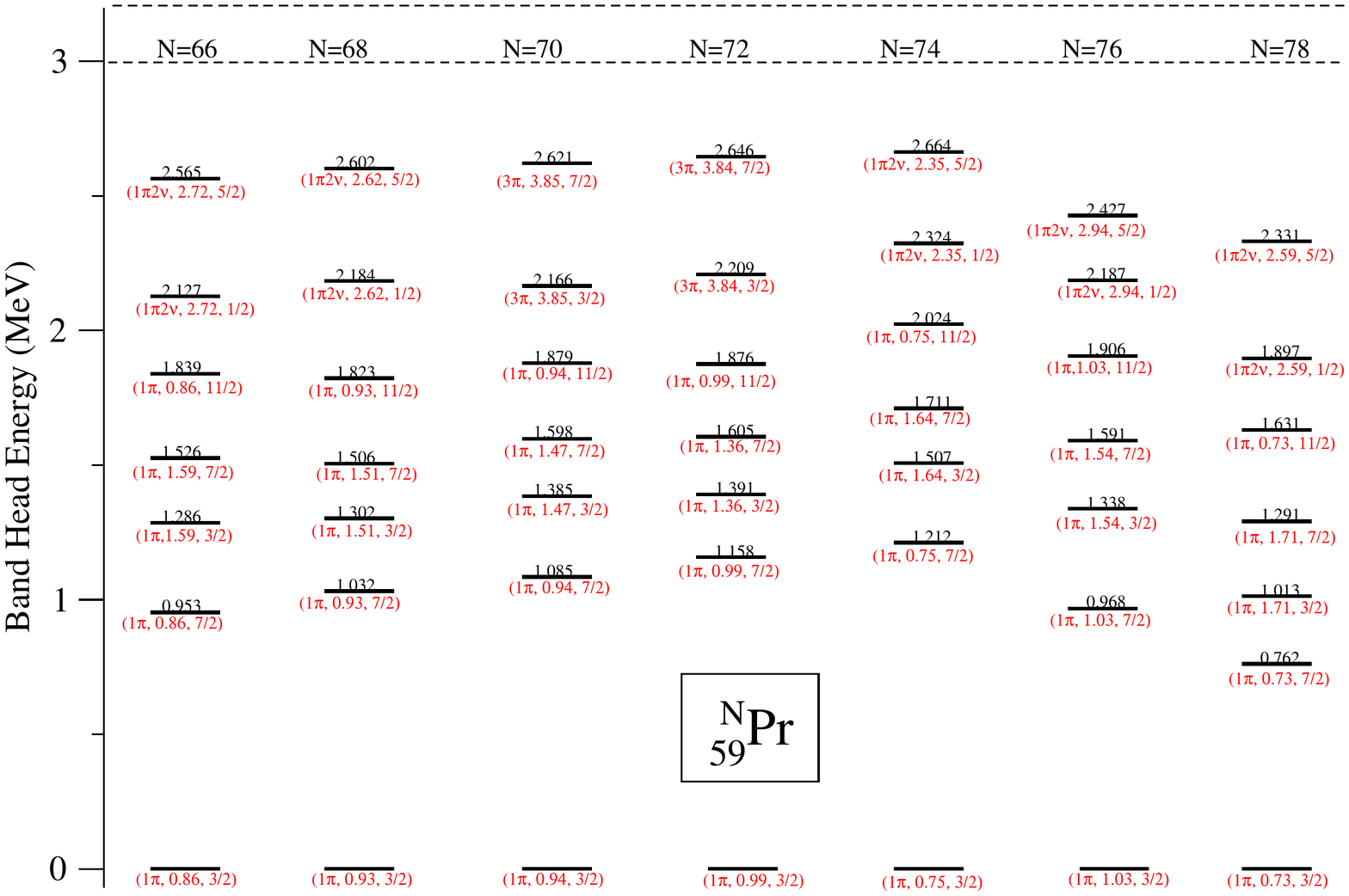} \caption{(Color
online)  TPSM band head energies after configuration mixing for
odd-proton $^{125-137}$Pr  isotopes. The dominant intrinsic
configuration is specified for each state. } \label{bhe1}
\end{figure*}
\begin{figure*}[htb]
\vspace{1cm}
 {\includegraphics[totalheight=17cm]{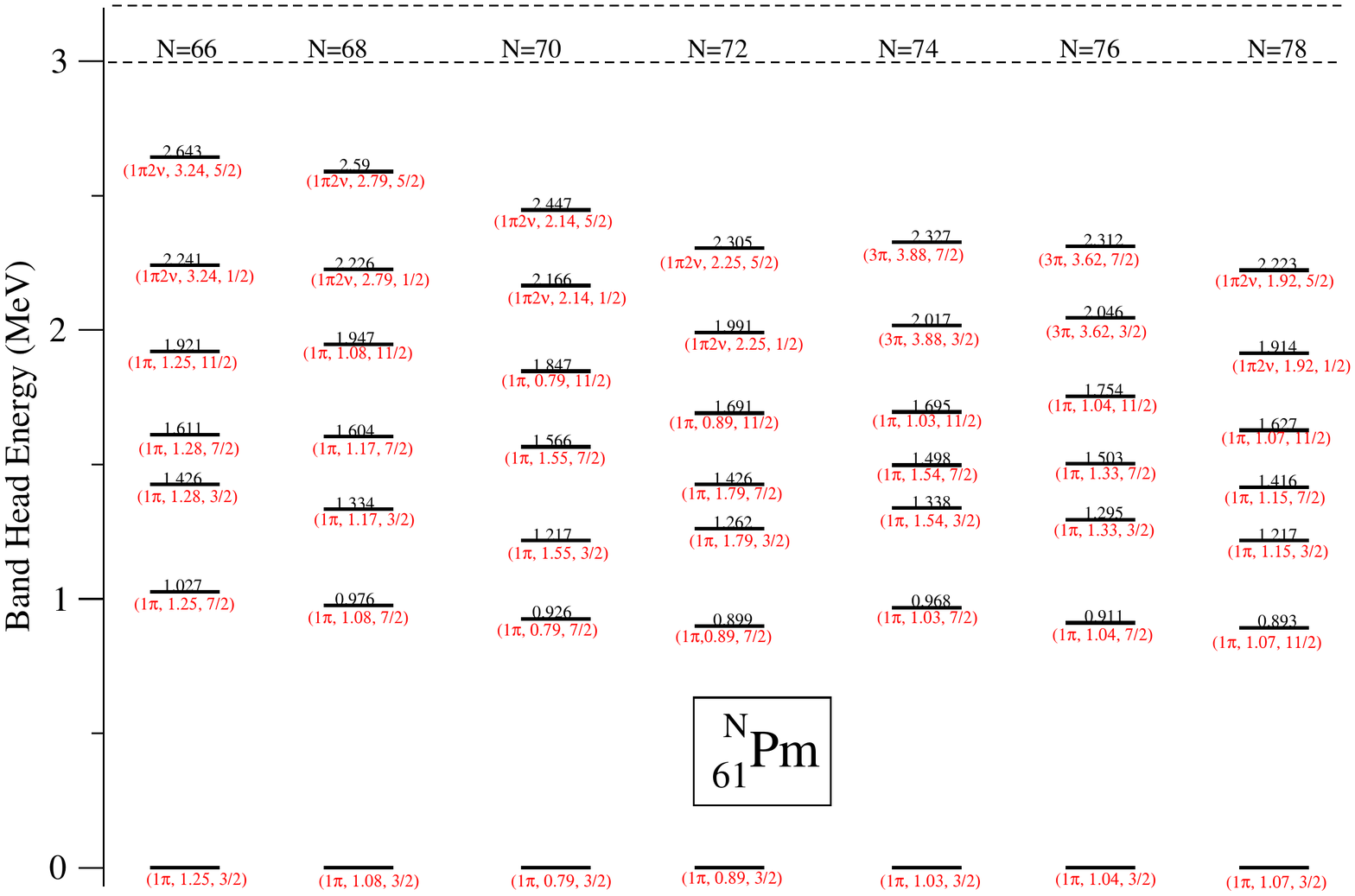}} \caption{(Color
online)  TPSM band head energies after configuration mixing for
odd-proton $^{127-139}$Pm  isotopes. The dominant intrinsic
configuration is specified for each state.} \label{bhe2}
\end{figure*}


\begin{figure*}[htb]
\vspace{0cm}
 {\includegraphics[totalheight=16cm]{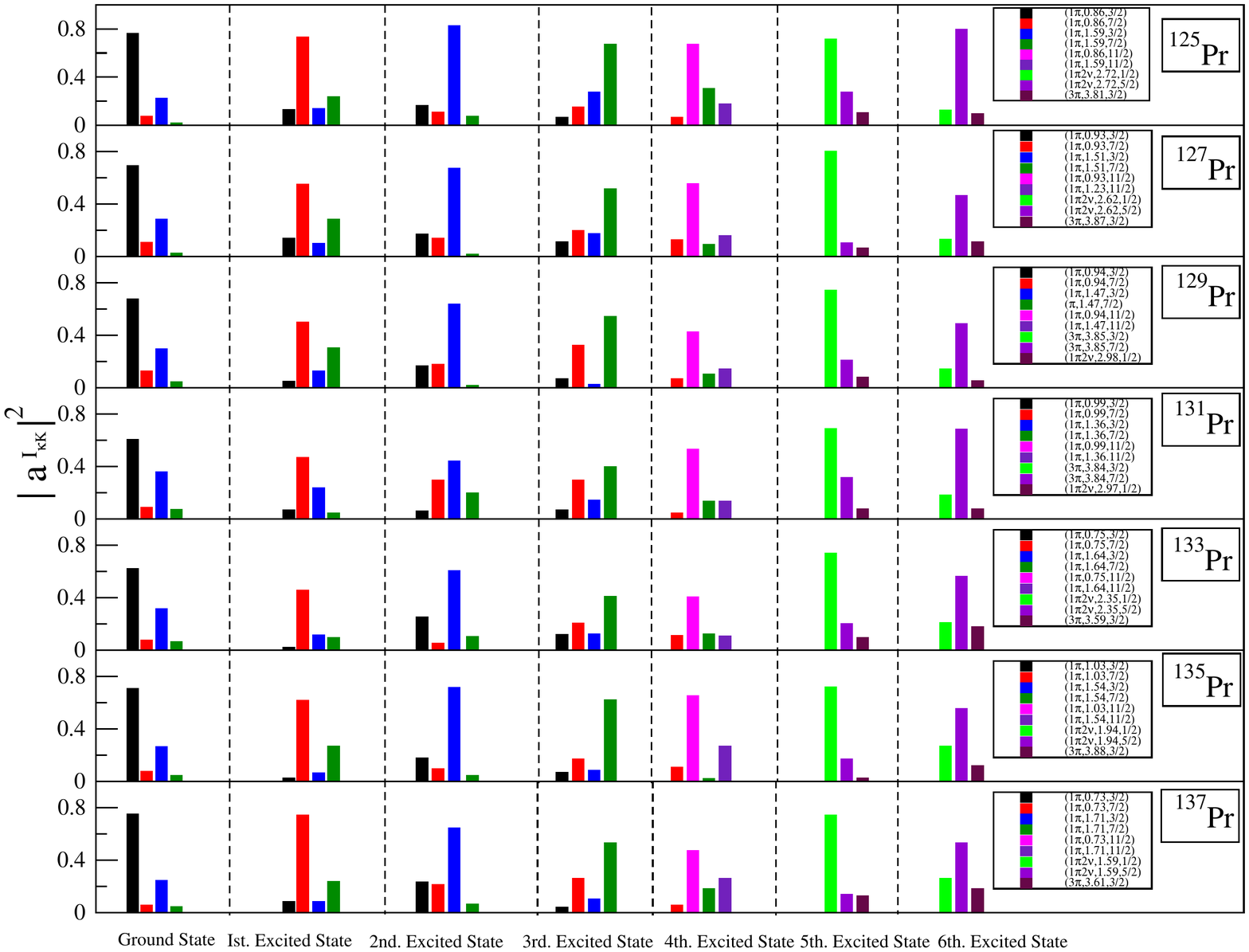}} \caption{(Color
online)   Dominant probability contributions of various projected configurations in the wave
functions of the band head structures shown in Fig.~\ref{bhe1}.} \label{wf1}
\end{figure*}
\begin{figure*}[htb]
\vspace{0cm}
 {\includegraphics[totalheight=16cm]{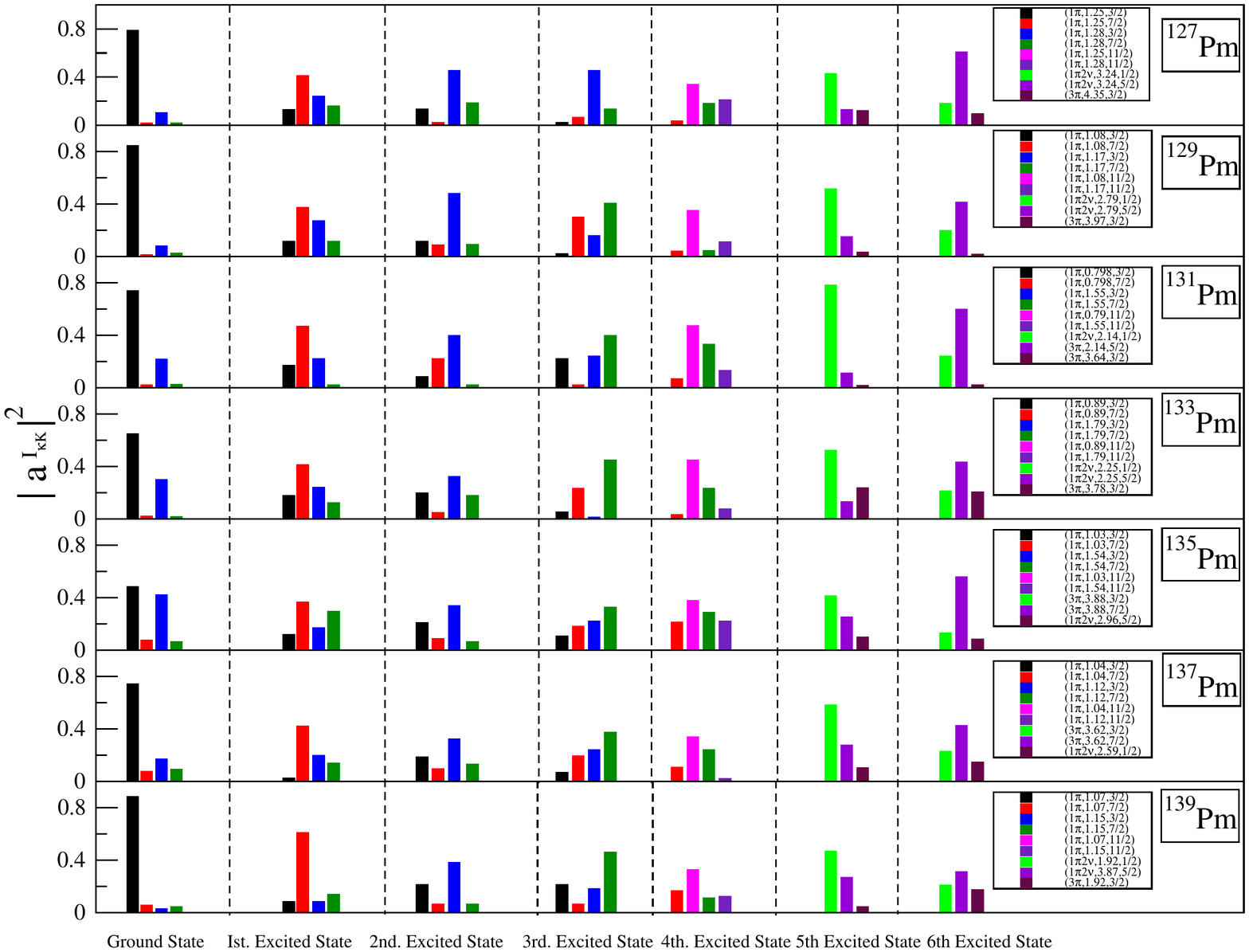}} \caption{(Color
online) Dominant probability contributions of various projected configurations in the wave
functions of the band head structures shown in Fig.~\ref{bhe2}. } \label{wf2} 
\end{figure*}
\begin{figure*}[htb]
\vspace{0cm}
\includegraphics[totalheight=16cm]{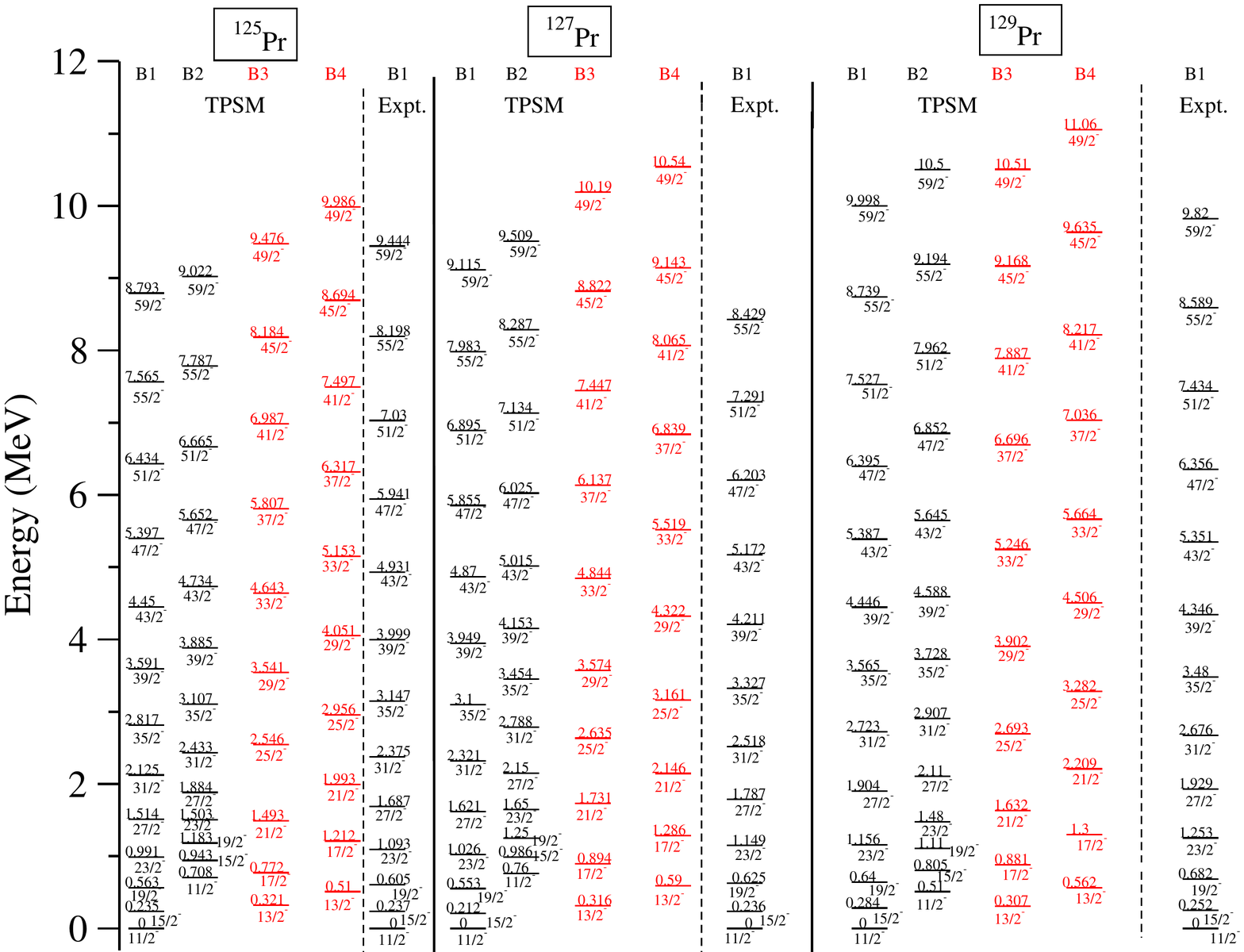} \caption{(Color
online) TPSM energies for the lowest two bands after configuration
mixing are plotted along
with the available experimental data for  $^{125,127,129}$Pr  isotopes. Data is
taken from \cite{Pr2529}.} \label{expe1}
\end{figure*}
\begin{figure*}[htb]
\vspace{0cm}
\includegraphics[totalheight=15cm]{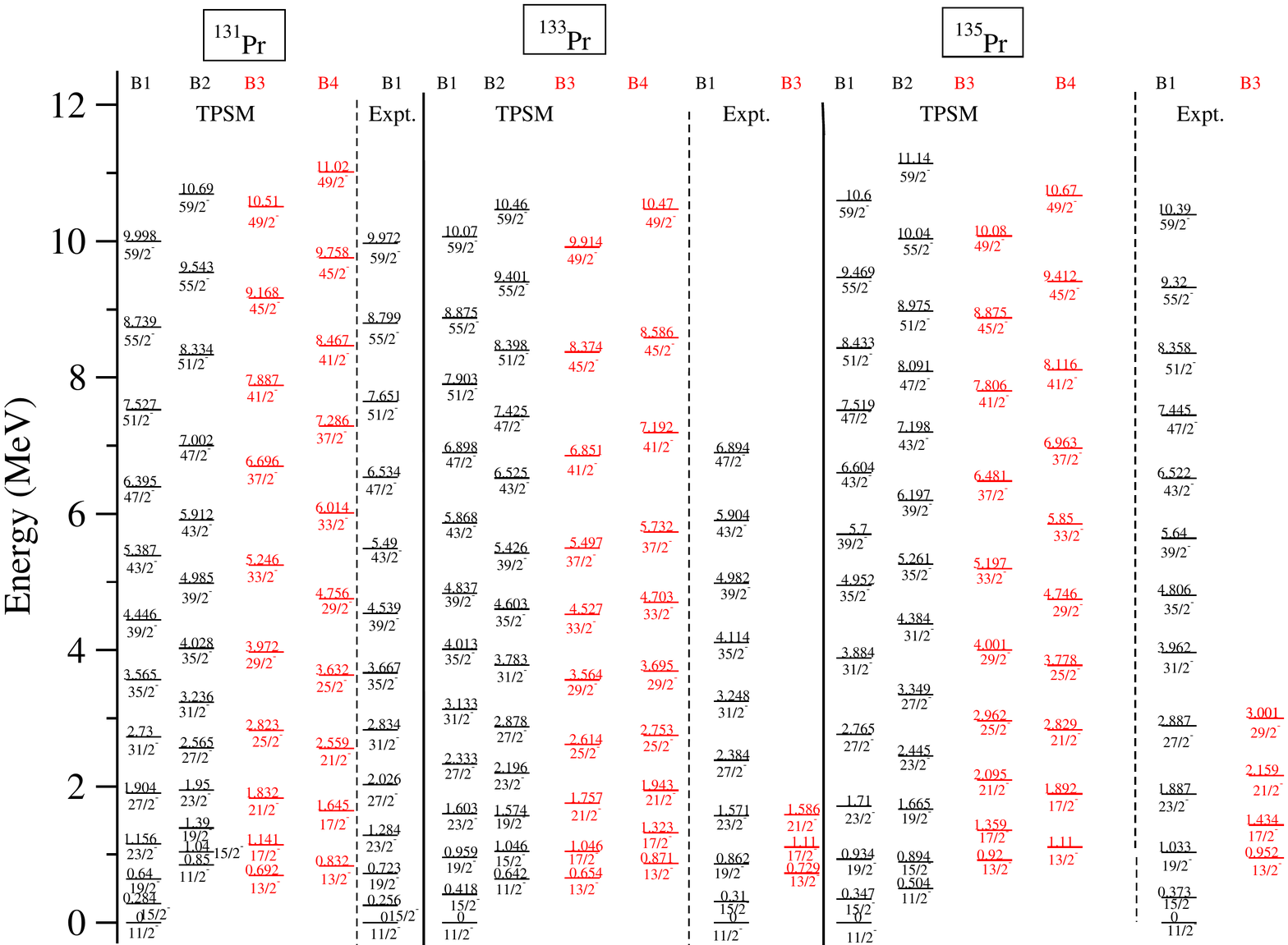} \caption{(Color
online)  TPSM energies for the lowest two bands after configuration
mixing are plotted along
with the available experimental data for  $^{131-135}$Pr  isotopes. Data  is
taken from \cite{Pr33,Pr35}.} \label{expe11}
\end{figure*}

\begin{figure}[htb]
 \centerline{\includegraphics[trim=0cm 0cm 0cm
0cm,width=0.6\textwidth,clip]{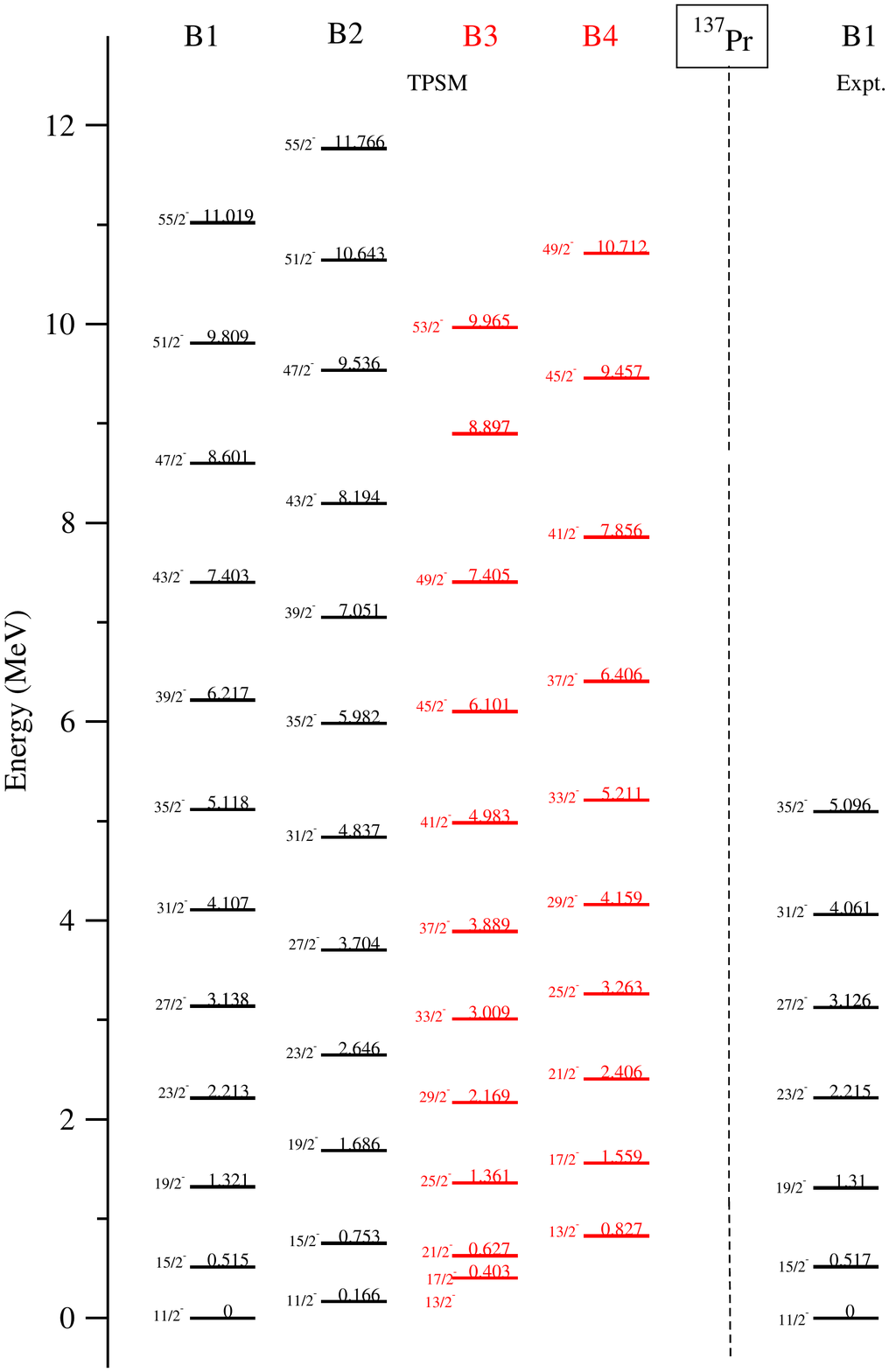}} \caption{(Color
online) TPSM energies  for the lowest two bands after configuration mixing
are plotted along with the available experimental data for  $^{137}$Pr  isotope. Data is
taken from \cite{Pr37}.} \label{expe2}
\end{figure}
\begin{figure*}[htb]
\vspace{0cm}
\includegraphics[totalheight=14cm]{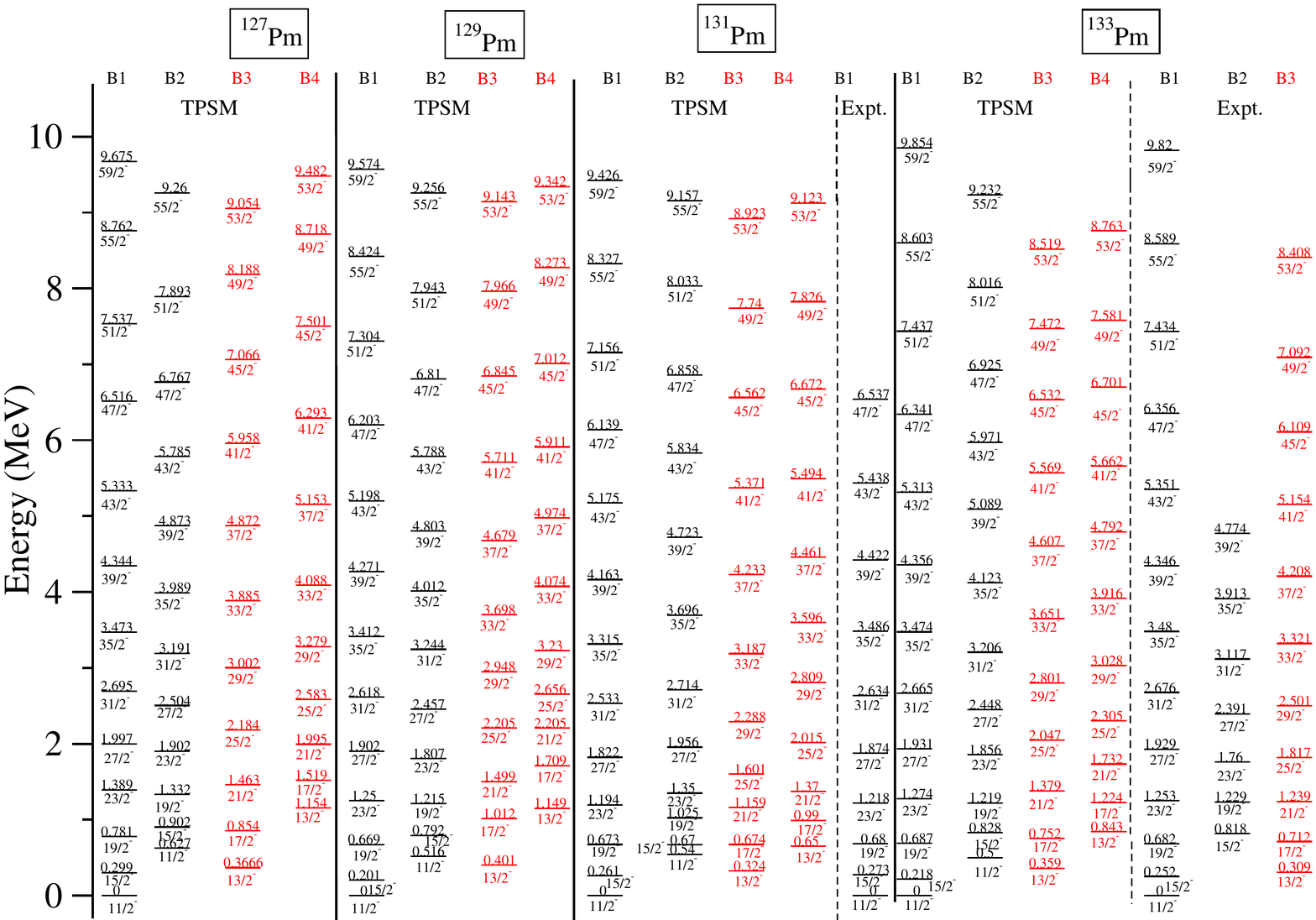} \caption{(Color
online)  TPSM energies for the lowest two bands after configuration
mixing are plotted along 
with the available experimental data for  $^{127-133}$Pm  isotopes. Data  is
taken from \cite{cm98}.} \label{expe101}
\end{figure*}
\begin{figure*}[htb]
\vspace{0cm}
\includegraphics[totalheight=16cm]{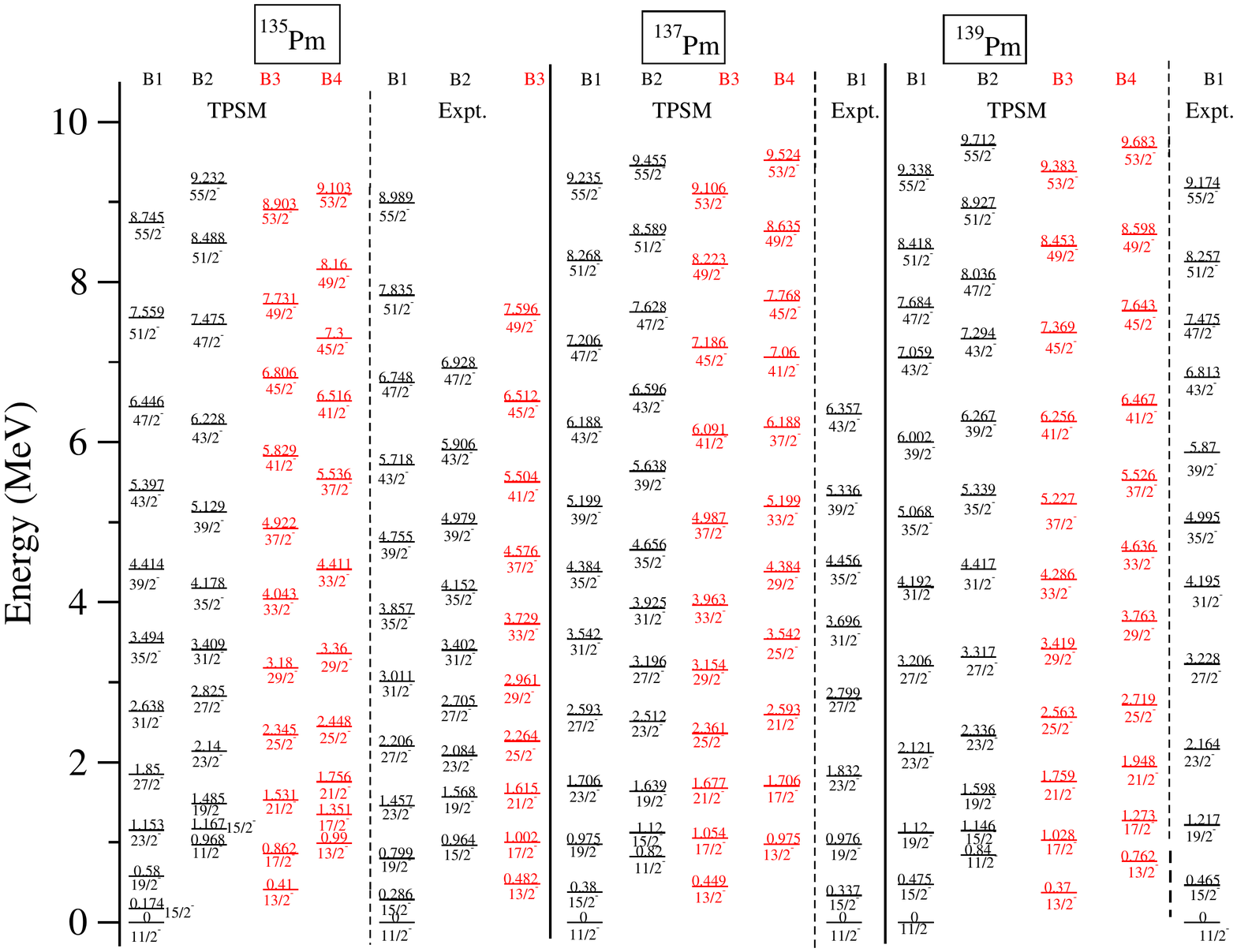} \caption{(Color
online) TPSM energies for the lowest two bands after configuration mixing
are plotted along with the available experimental data for  $^{135-139}$Pm  isotopes. Data is
taken from \cite{ad12,139pm}.} \label{expe22}
\end{figure*}

\begin{figure}[htb]
 \centerline{\includegraphics[trim=0cm 0cm 0cm
0cm,width=0.58\textwidth,clip]{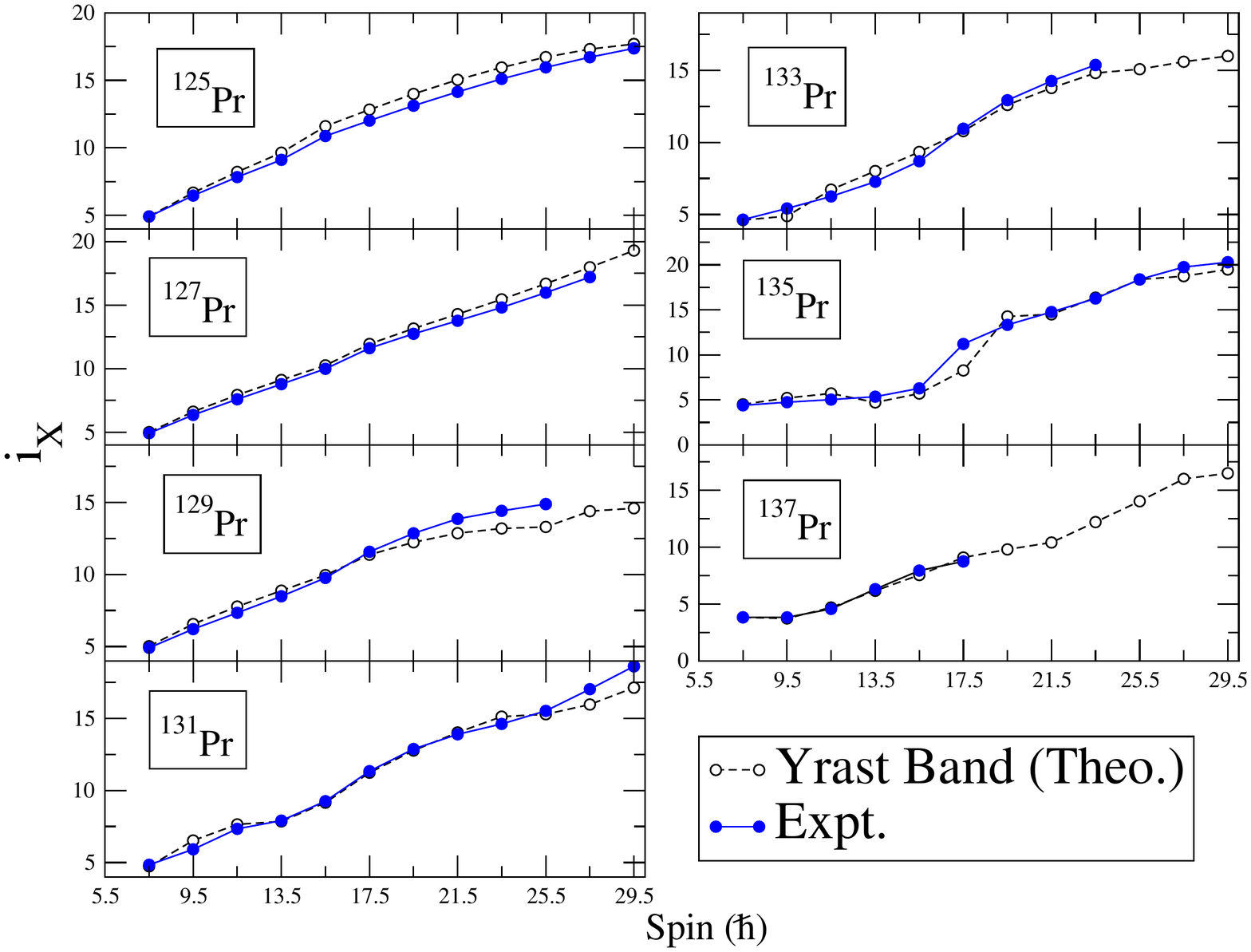}}
\caption{(Color online) Comparison of the aligned angular
 momentum,  $i_x=I_x(\omega)-I_{x,ref}(\omega)$, where $\hbar\omega=\frac{E_{\gamma}}{I_x^i(\omega)-I_x^f(\omega)}$,  $I_x(\omega)= \sqrt{I(I+1)-K^2}$  and $I_{x,ref}(\omega)=\omega(J_0+\omega^{2}J_1)$. The reference band Harris parameters used are $J_0$=23 and $J_1$=90, obtained from the measured energy levels
 as well as those calculated
from the TPSM results, for $^{125-137}$Pr nuclei. }\label{ali1}
\end{figure}
\begin{figure}[htb]
 \centerline{\includegraphics[trim=0cm 0cm 0cm
0cm,width=0.58\textwidth,clip]{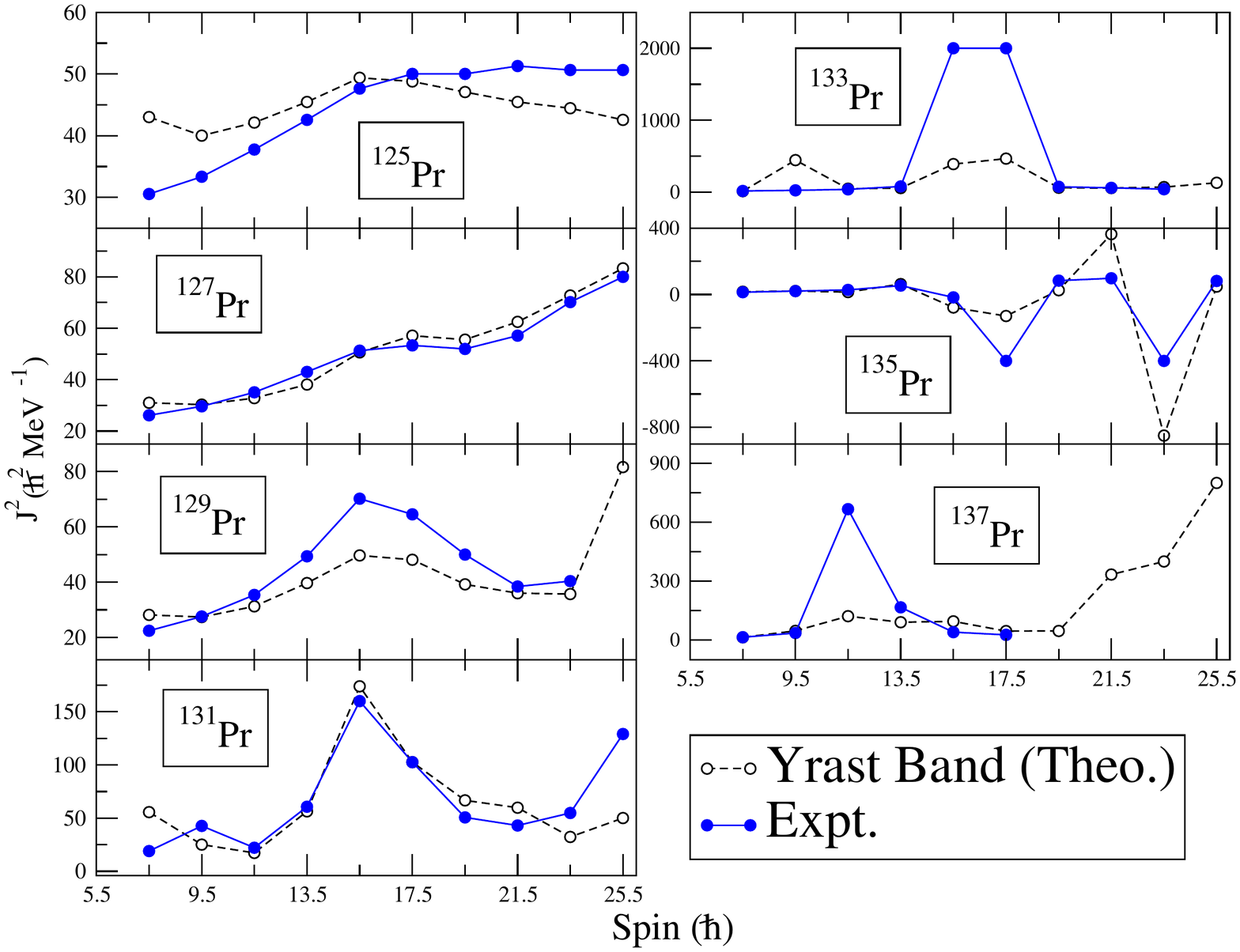}}
\caption{(Color online) Comparison between experimental and calculated
  dynamic moment of inertia,  J$^{(2)} =  \frac{4}{E_{\gamma}(I)-E_{\gamma}(I-2)}$,   
of the yrast band for $^{125-137}$Pr isotopes. }\label{ali2}
\end{figure}
\begin{figure}[htb]
 \centerline{\includegraphics[trim=0cm 0cm 0cm
0cm,width=0.58\textwidth,clip]{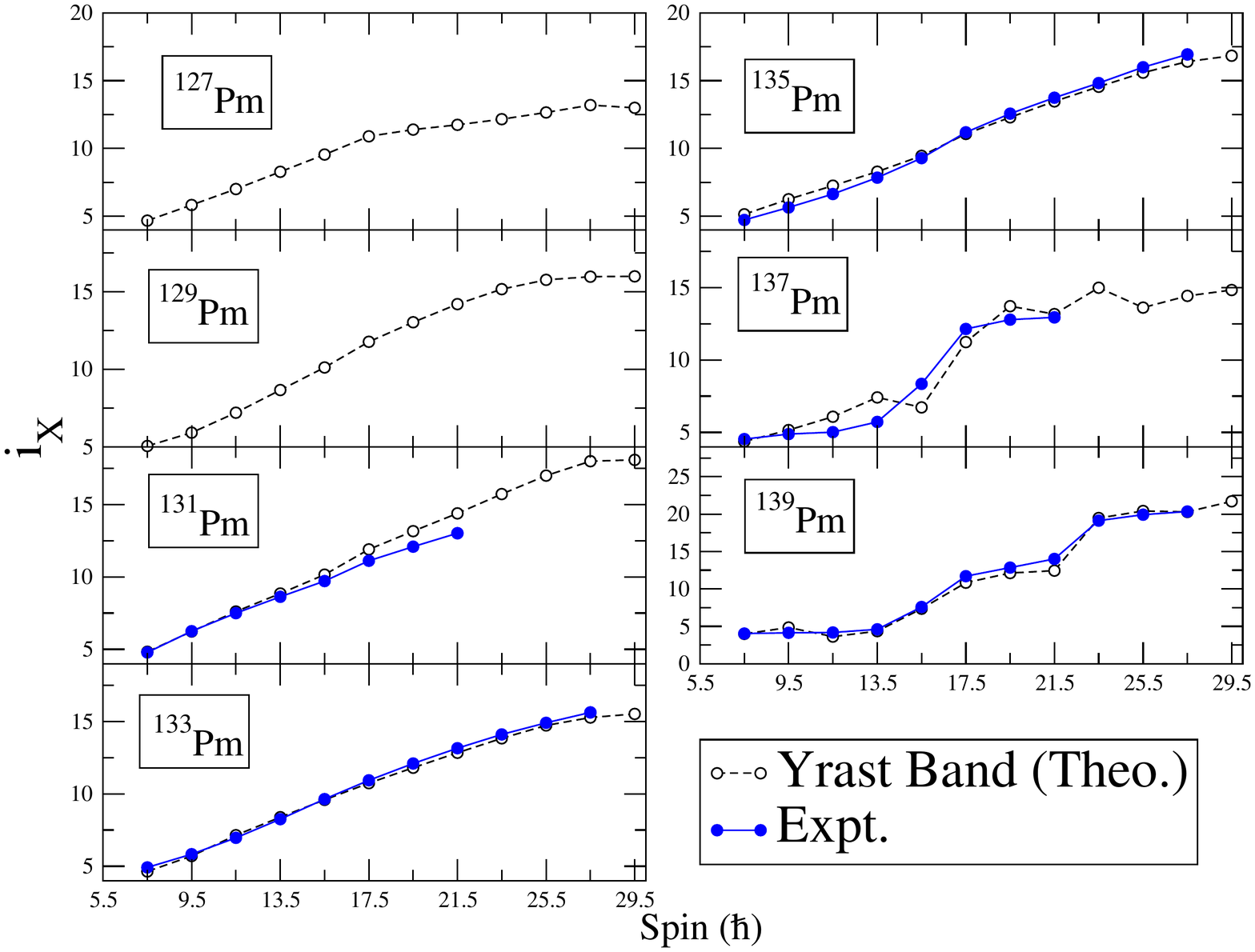}}
\caption{(Color online) Comparison of the measured and calculated aligned angular
 momentum ($i_x$)  for $^{127-139}$Pm nuclei. }\label{ali3}
\end{figure}

\begin{figure}[htb]
 \centerline{\includegraphics[trim=0cm 0cm 0cm
0cm,width=0.55\textwidth,clip]{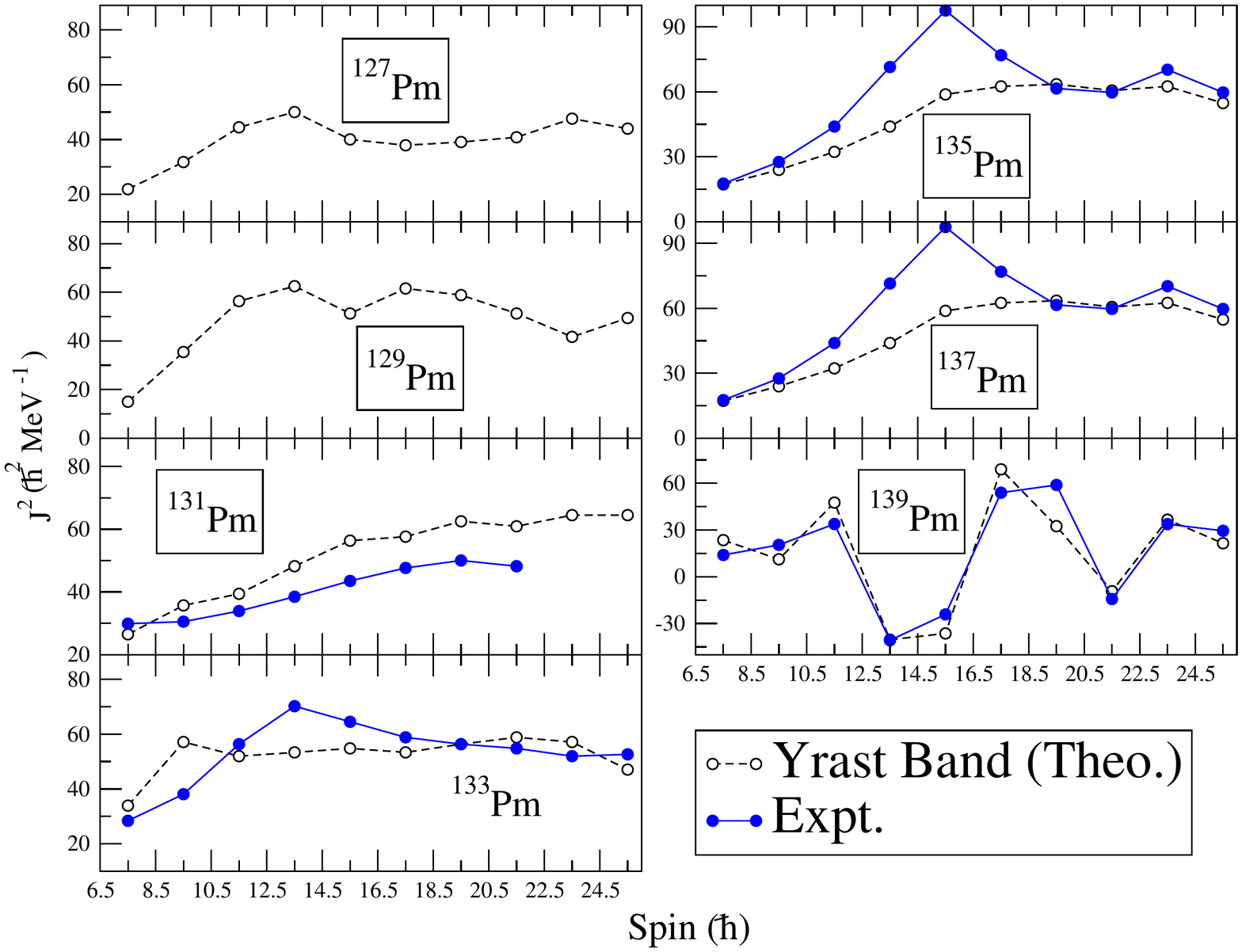}}
\caption{(Color online) Comparison between experimental and calculated
  dynamic moment of inertia 
(J$^{(2)}$)  of the yrast band for $^{127-139}$Pm isotopes. }\label{ali4}
\end{figure}
\begin{figure}[htb]
 \centerline{\includegraphics[trim=0cm 0cm 0cm
0cm,width=0.52\textwidth,clip]{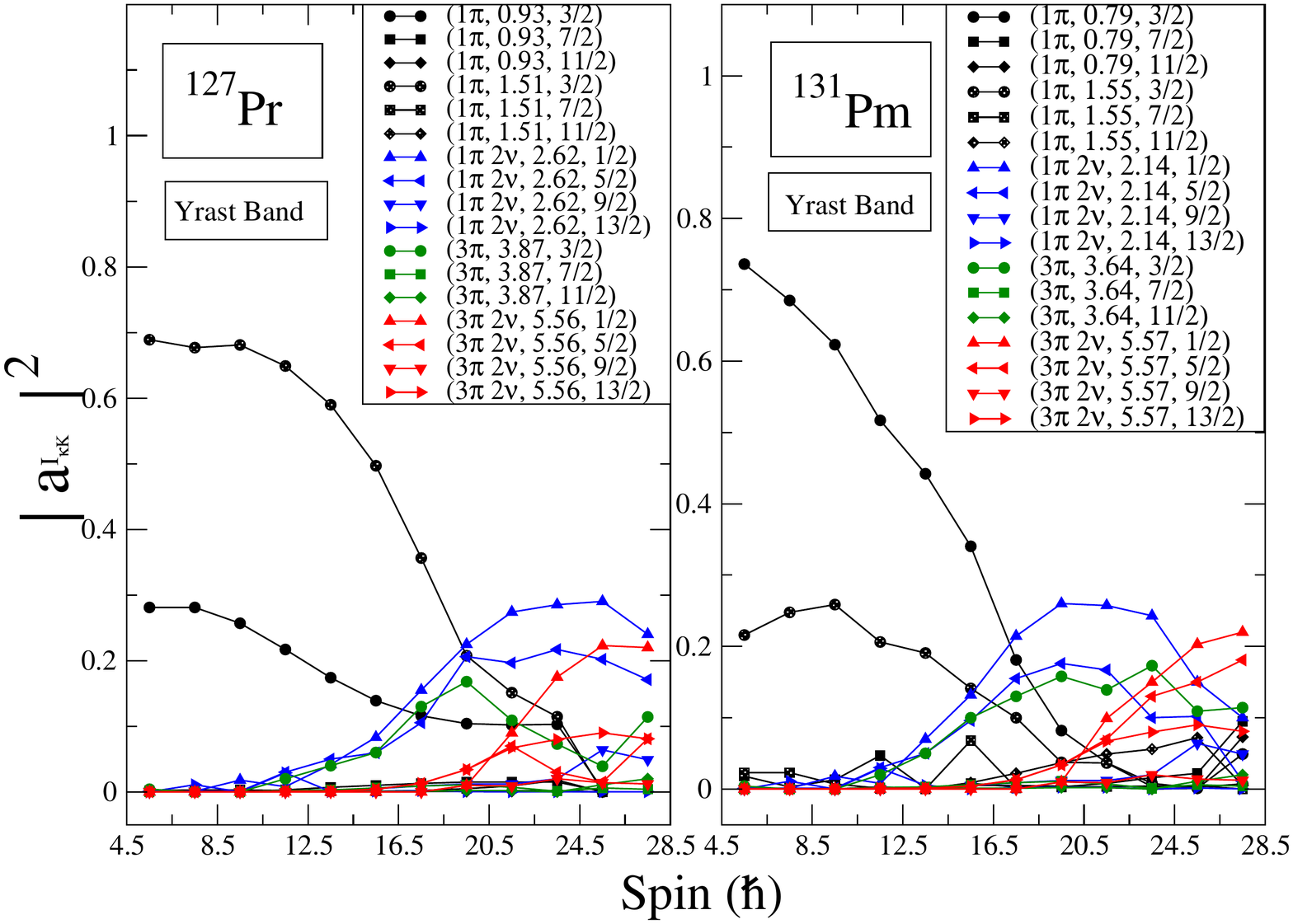}}
\caption{(Color online) Probability of various projected
  K-configurations in the wave
functions of the yrast band after
diagonalization for 
  $^{127}$Pr and  $^{131}$Pm isotopes.} \label{wf3}
\end{figure}

It has been shown
using the TPSM approach that each quasiparticle state has a $\gamma$-
band built on it \cite{scr} and the second excited s-band, as a matter of fact, is a $\gamma$-band built on the
two-quasiparticle state. As the intrinsic structures of the
two-quasiparticle band and the $\gamma$-band built on it is same, the
predicted g-factors for two bands should be similar. This explained 
as to why both the observed s-bands have either positive or negative
g-factor values \cite{sj17,zemel}. 

As Cerium and Neodymium are even-even cores of 
odd-mass Praseodymium and Promethium nuclei, it is expected that
these odd-mass nuclei should also depict $\gamma$-bands built on
quasiparticle states. $\gamma$-bands in some odd-mass nuclei have
already been identified \cite{sh10}. To delineate $\gamma$-bands in
high-spin band structures of  odd-mass Pr- and
Pm- isotopes is one of the objectives
of the present work. The manuscript is organised in the following 
manner. In the next section, the extended TPSM approach is briefly
presented. In section III, TPSM results obtained for Pr- and Pm- isotopes
are compared with the experimental data, where ever
available. Finally, the summary and conclusions obtained in the
present study are outlined in section IV.

\section{Triaxial Projected Shell Model Approach}
The inclusion of multi-quasiparticle basis space in TPSM approach has
made it feasible to study not only the ground-state properties, but
also the high-spin band structures in deformed and transitional nuclei
\cite{sj17,scr,ns19}. Using the TPSM approach, odd-proton systems have
been studied earlier with the model space of one-proton and one-proton coupled to
two-neutron quasiparticle states. However, in order to investigate the
high-spin spectroscopy of these systems, the basis space needs to be
extended by including proton aligning configurations, in addition to
the neutron states. In the present work, the extended basis space has
been implemented and the complete basis space in the generalized approach is given by $:$
\begin{equation}
\begin{array}{r}
~~\hat P^I_{MK}~ a^\dagger_{\pi_1} \ack\Phi\ket;\\
~~\hat P^I_{MK}~a^\dagger_{\pi_1}a^\dagger_{\nu_1}a^\dagger_{\nu_2}  \ack\Phi\ket;\\
~~\hat P^I_{MK}~a^\dagger_{\pi_1}a^\dagger_{\pi_2}a^\dagger_{\pi_3}  \ack\Phi\ket;\\
~~\hat P^I_{MK}~a^\dagger_{\pi_1}a^\dagger_{\pi_2} a^\dagger_{\pi_3}a^\dagger_{\nu_1} a^\dagger_{\nu_2} \ack\Phi\ket,
\label{intrinsic}
\end{array}
\end{equation}
where $\ack\Phi\ket$ is the triaxially-deformed quasiparticle vacuum state. $P^I_{MK}$ is the three-dimensional
angular-momentum-projection operator given by \cite{RS80} $:$ 
\begin{equation}
\hat P ^{I}_{MK}= \frac{2I+1}{8\pi^2}\int d\Omega\, D^{I}_{MK}
(\Omega)\,\hat R(\Omega),
\label{Anproj}
\end{equation}
with the rotation operator 
\begin{equation}
\hat R(\Omega)= e^{-i\alpha \hat J_z}e^{-i\beta \hat J_y}
e^{-i\gamma \hat J_z}.\label{rotop}
\end{equation}
Here, $''\Omega''$ represents the set of Euler angles 
($\alpha, \gamma = [0,2\pi],\, \beta= [0, \pi]$) and  
$\hat{J}^{,}s$ are the angular-momentum operators. The
angular-momentum projection operator in Eq.~(\ref{Anproj}) not only
projects out the good angular-momentum, but also states 
having good $K$-values by specifying a value for
$K$ in the rotational matrix, $"D"$ in Eq.~(\ref{Anproj}).

The  constructed projected basis of Eq.~(\ref{intrinsic})
is then used to diagonalise the shell model Hamiltonian, consisting of
the harmonic oscillator single-particle Hamiltonian and a  
residual two-body interaction comprising of
quadrupole-quadrupole, monopole pairing  and quadrupole pairing terms. 
These terms represent specific correlations which are considered to be
essential to describe the low-energy nuclear phenomena \cite{BarangerKumar}. The Hamiltonian has the following form $:$
\begin{eqnarray}
\hat H =  \hat H_0 -   {1 \over 2} \chi \sum_\mu \hat Q^\dagger_\mu
\hat Q^{}_\mu - G_M \hat P^\dagger \hat P - G_Q \sum_\mu \hat
P^\dagger_\mu\hat P^{}_\mu . \label{hamham}
\end{eqnarray}
In the above equation, $\hat H_0$ is the spherical single-particle
part of the  Nilsson potential \cite{Ni69}. 
The QQ-force strength, $\chi$, in Eq. (\ref{hamham}) is related to
the quadrupole deformation $\epsilon$ as a result of the
self-consistent HFB condition and the relation is given by
\cite{KY95}:
\begin{equation}
\chi_{\tau\tau'} =
{{{2\over3}\epsilon\hbar\omega_\tau\hbar\omega_{\tau'}}\over
{\hbar\omega_n\left<\hat Q_0\right>_n+\hbar\omega_p\left<\hat
Q_0\right>_p}},\label{chi}
\end{equation}
where $\omega_\tau = \omega_0 a_\tau$, with $\hbar\omega_0=41.4678
A^{-{1\over 3}}$ MeV, and the isospin-dependence factor $a_\tau$ is
defined as
\begin{equation}
a_\tau = \left[ 1 \pm {{N-Z}\over A}\right]^{1\over 3},\nonumber
\end{equation}
with $+$ $(-)$ for $\tau =$ neutron (proton). The harmonic
oscillation parameter is given by $b^2_\tau=b^2_0/a_\tau$ with
$b^2_0=\hbar/{(m\omega_0)}=A^{1\over 3}$ fm$^2$. The monopole pairing strength $G_M$ (in MeV)
is of the standard form
\begin{eqnarray}
G_M = {{G_1 \mp G_2{{N-Z}\over A}}\over A}, 
 \label{pairing}
\end{eqnarray}
where the minus (plus) sign applies to neutrons (protons). In the present calculation, we choose  $G_1$ and $G_2$ such that the 
calculated 
gap parameters reproduce the experimental mass  differences. This choice of $G_M$ is appropriate for the single-particle space employed 
in the present calculation, where three major oscillator shells are used for each
type of nucleons (N$=3,4,5$ major shells for both neutrons and protons). The quadrupole pairing strength $G_Q$ is assumed to be 
proportional to $G_M$, the proportionality constant being fixed as usual to be 0.16.  These interaction strengths, although not
exactly the same, are consistent with those used earlier in the TPSM calculations \cite{JG12,bh14,sh10}.\\

Using the angular-momentum projected states as the basis, the shell
model Hamiltonian of Eq.~(\ref{hamham}) is diagonalized 
following the Hill-Wheeler approach \cite{KY95}. The generalized eigen-value
equation is given by 

\begin{equation}
 \sum_{\kappa^{'}K^{'}}\{\mathcal{H}_{\kappa K \kappa^{'}K^{'}}^{I}-E\mathcal{N}_{\kappa K 
\kappa^{'}K^{'}}^{I}\}f^{I}_{\kappa^{'} K^{'}}=0, \label{a15}
\end{equation}
 where the Hamiltonian and norm kernels are given by
 \begin{eqnarray*}
 && \mathcal{H}_{\kappa K \kappa^{'}K^{'}}^{I} = \langle \Phi_{\kappa}|\hat H\hat 
P^{I}_{KK^{'}}|\Phi_{\kappa^{'}}\rangle ,\\
&&\mathcal{N}_{\kappa K \kappa^{'}K^{'}}^{I}= \langle \Phi_{\kappa}|\hat P^{I}_{KK^{'}}|\Phi_{\kappa^{'}}\rangle .
 \end{eqnarray*}
 The Hill-Wheeler wavefunction is given by
\begin{equation}
  |\psi_{IM}\ket = \sum_{\kappa K} a_{\kappa K}^{I}\hat {P}_{MK}^{I} |\Phi_{\kappa} \ket,\label{Anprojaa}
 \end{equation}
where $ a_{\kappa K}^{I}$ are the variational coefficients, and the
index  $^{``}\kappa^{``}$ designates the basis states of Eq. (\ref{intrinsic}).

\section{Results and Discussion}

TPSM calculations have been performed for odd-mass $^{125-137}$Pr and $^{127-139}$Pm
isotopes using the axial and non-axial deformations listed in 
Table 1. These deformation values have been adopted from the earlier
studies performed for these nuclei \cite{fs21,ysm,MN95}. The intrinsic states
obtained from the solution of the triaxial Nilsson potential with
these deformation parameters are projected onto good angular-momentum states as
discussed in the previous section. For each system about 40 to 50
intrinsic states are selected around the Fermi surface for which
the angular-momentum projection is performed. The angular-momentum
projected states, which are close to the yrast line, are depicted in
Fig.~\ref{bd1} for $^{125}$Pr.  We would like to mention that all the
projected states near the fermi surface are employed in the final diagonalization of the
shell model Hamiltonian, but for clarity only projected states that 
are close to the yrast line are shown in Fig.~\ref{bd1}. This diagram, what 
is referred to as the band diagram, is quite instructive as it reveals
the intrinsic structures of the observed band structures \cite{KY95}.

The lowest projected band in Fig.~\ref{bd1} originates from the
one-quasiproton state, having K $=3/2$, with the quasiparticle
energy of 0.86 MeV. Although the triaxial quasiparticle state does
not have a well-defined angular-momentum projection quantum number,
but the three-dimensional projection operator, not only projects out the angular-momentum
quantum number, but also its projection along the intrinsic z-axis,
what is referred to as the K-quantum number in the literature
\cite{lamme,boeker}.
The K-value specified in all the diagrams in the present work refers to this projected 
quantum number. The two signature branches for low-K bands are
shown separately as the splitting between the two states is quite
large for these configurations. The signature splitting for the 
lowest K $=3/2$ band increases as expected with
increasing angular-momentum. It is noted from Fig.~\ref{bd1} that
three-quasiparticle band comprised of one-proton and two-neutron aligned
configuration, having K= $1/2$, crosses the ground-state band and becomes yrast
at I $= 41/2$. This crossing is between the $\alpha=-1/2$ states of the
two bands, and the $\alpha=+1/2$ branch is quite high in excitation
energy as compared to its signature partner band.

What is interesting to note from Fig.~\ref{bd1} is that 
above the band crossing, the $\alpha=+1/2$ states of the yrast band
originate from the $\gamma$-band built on the three quasiparticle state. In
the TPSM analysis, $\gamma$-bands are built on each quasiparticle
state \cite{scr} and apart from the $\gamma$-band based on the ground-state,
$\gamma$-bands built on two-quasiparticle states have been identified 
in several even-even nuclei \cite{156dy,zemel,sj17,kum14}. These bands have $K=K_0+2$, where $K_0$
is the K-quantum number of the parent band. In the present work, 
$\gamma$-band based on the ground-state, K $=3/2$, has K $=7/2$ and is located
at an excitation energy of $\sim 1.0$ MeV from the ground-state band at
I $=11/2$. 
The $\gamma$-band built on the aligned three quasiparticle state with
K $=1/2$ has K $=5/2$ and is located at an excitation energy of $\sim
2.5 $ MeV
at I $=11/2$. This band is noted to cross the
$\gamma$-band based on the ground-state band at I $=35/2$ and
becomes the lowest band for $\alpha=+1/2$ signature branch. The
reason that it becomes lowest is because it is built on the
three-quasiparticle state with K $=5/2$ and depicts
less signature splitting as compared to its parent band. It is also
noted from Fig.~\ref{bd1} that a five-quasiparticle state, having both 
two-protons and two-neutrons aligned, with K $=1/2$, crosses the 
three-quasiparticle state at a higher spin, I $=59/2$.

The band diagrams for other studied Pr-isotopes are similar to that
of $^{125}$Pr, except that nature of the band crossing changes with increasing neutron
number. In Fig.~\ref{bd3}, only the band crossing portion of the band
diagrams is depicted for the Praseodymium isotopes ranging from
A=127 to 137.  The first band crossing in $^{127}$Pr is again due to
the alignment of two-neutrons, but occurs at I $=39/2$, which 
is slightly lower as compared to that of $^{125}$Pr. For $^{129}$Pr, the
nature of the first band crossing has changed and is now due to the
alignment of two-protons rather than of two-neutrons as was for
the  earlier two cases. The three-proton configuration having K $=3/2$ 
becomes lower than one-quasiparticle ground-state band at
I $=35/2$. For $^{131}$Pr, the band crossing occurs at I $=35/2$ as 
for $^{129}$Pr, but for other studied isotopes, it is observed at a
higher angular-momentum of I $=43/2$. These band crossing features
shall be discussed in detail later when comparing the alignment and
moment of inertia obtained from the TPSM results with those deduced
from the experimental data.

The band structures for $^{127}$Pm are displayed in Fig.~\ref{bd2} and again
only configurations important to describe the near yrast spectroscopy
are plotted. The ground-state band, having K $=3/2$, is built on the one-quasiproton 
Nilsson state with energy of $0.93$ MeV.  The $\gamma$-band based on
the ground-state band, having K $=7/2$ lies at an excitation energy of
$\sim 1$ MeV  for I $=11/2$. It is observed from the figure that
three-quasiparticle state having K $=1/2$ with one-proton coupled to two-aligned
neutrons, crosses the ground-state band at I $=43/2$. It is also 
noted from the figure that $\gamma$-band built on
the three-quasiparticle state, having K $=5/2$, also crosses the
ground-state band at I $=47/2$. This almost simultaneous crossing
will lead to forking of the ground-state band into two s-bands as is
known to occur in many even-even systems \cite{sj17, wyss1}. 

In $A \sim 130$ region, 
some even-even isotopes of Ba, Ce and Nd are known to have several
s-bands \cite{wyss1}. As the neutron and proton Fermi surfaces are close in energy 
for these isotopes, the forking of the ground-state band into two
s-bands is expected with one s-band having neutron character and the 
other originating from protons. However, this traditional picture
cannot explain the magnetic moment measurements for the band heads,
I $=10^+$ states, of the two s-bands with g-factors of both the
states having a neutron character in $^{134}$Ce \cite{zemel}. This long-standing
puzzle was addressed using the TPSM approach and it was shown
\cite{JG09} that the second s-band in $^{134}$Ce is a $\gamma$-band based on the
two-neutron aligned state and since the intrinsic configurations of
the two s-bands are same in this interpretation, the g-factors
of the two s-bands are expected to be similar \cite{sj17}. It was also
predicted that two s-bands observed in $^{136,138}$Nd nuclei should 
both have positive g-factors with the aligning particles being
protons \cite{sj17}. Further, $\gamma$-bands built on two-quasiparticle
states have been observed in $^{70}$Ge \cite{kum14} and $^{156}$Dy
\cite{156dy} nuclei. 

In the present work, we shall examine whether it would be feasible to 
identify the $\gamma$-bands built on quasiparticle states.
$\gamma$-band in odd-mass nuclei are quite rare and these bands built
on the ground-state have been been identified in
$^{103,105}$Nb \cite{sh10,hj13}, $^{107,109}$Tc \cite{gu10}, and very recently
in $^{155,157}$Dy nuclei \cite{snt20}. The problem is that in odd-mass
nuclei, $\gamma$- configurations compete with 
one-quasiparticle states and contain a strong admixtures from these
states. This shall be addressed later in the presentation of the band
head energies of various band structures after diagonalization of the
shell model Hamiltonian.

The band diagrams for other studied Pm-Isotopes are displayed in
Fig.~\ref{bd4}, depicting only the important band crossing regions. The 
band crossing for $^{129}$Pm and $^{131}$Pm isotopes occur at
I $=39/2$ and is due to the alignment of two-neutrons. For other 
Pr-isotopes, the band crossing occurs at lower angular-momenta
and is due to the alignment of two-protons. It is also noted from 
Fig.~\ref{bd1} that the five-quasiparticle state, which contains three-proton
plus two-neutron aligned configuration, crosses the
three-quasiparticle state at higher angular-momenta.
Therefore, the present
calculations predict that odd-Pm isotopes, studied in the present
work, should depict a second crossing at high-spin. 

The angular-momentum projected states, depicted in the band diagrams Figs. \ref{bd1}, \ref{bd3}, \ref{bd2} and \ref{bd4} and many more in the vicinity of the Fermi surface are employed to
diagonalize the shell model Hamiltonian of Eq. (\ref{hamham}). As already
mentioned that present approach is similar to the traditional
spherical shell model (SSM) approach with the exception that angular-momentum states,
projected from the deformed Nilsson configurations, are employed as
the basis states instead of the spherical configurations. The nuclei
studied in the present work are beyond the reach of the SSM approach
as the dimensionality of the spherical basis space becomes too prohibitive
to manage with the existing computational facilities. In the TPSM
approach, the optimal deformed basis states are choosen to describe
the properties of deformed systems and the number of basis states
required is quite minimal. In most of the studies, it has been
demonstrated that 40 to 50 basis states are sufficient to describe the
properties of deformed systems. The additional work required in the
TPSM approach is that the deformed basis need to be projected onto
states having good angular-momentum in order to diagonalize the
spherical shell model Hamiltonian of Eq. (\ref{hamham}).

In the studied Pr-isotopes, the lowest projected states after 
diagonalization are depicted in
Fig.~\ref{bhe1}  for the angular-momentum, I $=11/2$, which is the
ground-state for the negative parity bands observed for the studied isotopes.
The states in Fig.~\ref{bhe1} are labelled with the projected
intrinsic state that is 
most significant in the wavefunction. It is noted that ground-state for
all the studied isotopes originates from the one-quasiparticle state
having K $=3/2$. The $\gamma$-band based on the ground-state,
having K $=7/2$, is located at about $1$ MeV excitation energy from the
ground-state in all the isotopes. The main problem to identify them in 
odd-mass systems is that they are mixed with the single-particle
states as is evident from the figure that there are several
single-quasiparticle states, which are in the vicinity of the
$\gamma$-bands. This figure also displays, three-quasiparticle states and the
$\gamma$-bands built on them. These three-quasiparticle states become 
favoured in energy at high-spin and cross the
ground-state band as illustrated in the band diagrams,
Figs. \ref{bd1}, \ref{bd3}, \ref{bd2} and \ref{bd4}.
It is also noted from Fig.~\ref{bhe1} that these three-quasiparticle states
become lower in energy for $^{135}$Pr and $^{137}$Pr isotopes and it might be
feasible to populate the low-spin members of these states. In particular,
the most interesting prediction is the possibility of 
observing almost two identical three-quasiparticle bands, 
one the normal three-quasiparticle band having K $=1/2$
and the other the $\gamma$-band, having K $=5/2$, based on the
three quasiparticle state.  These
two states should have similar electromagnetic properties, like
g-factors, since they originate from the same intrinsic quasiparticle
configuration. 

The band head energies for the studied Pm-isotopes are displayed in
Fig.~\ref{bhe2} and have a similar pattern as that for the Pr-isotopes,
shown in Fig.~\ref{bhe1}. The only difference between the two figures is
that the band head energies for the three-quasiparticle band
structures is slightly lower for the Pm-isotopes. In particular,  for 
$^{139}$Pm, the three quasiparticle state and the $\gamma$-band 
based on this state is quite low in energy and is the  best candidate 
for which these structures could be identified in the future
experimental studies.

The dominant components in the wavefunctions of the above 
discussed band head states are depicted in Figs.~\ref{wf1} and \ref{wf2}
for I $=11/2$. It is observed from these figures that all the states
are mixed, even the ground-state band head has small admixtures 
from the other one-quasiparticle states and also from the
$\gamma$-band. We would like to remind the reader that projection from a triaxial 
intrinsic state give rise to several bands having different values of
the K-quantum number. The shell model Hamitonian is diagonalized with
all these projected states that results into mixing among
them. $\gamma$-band, which happens to be the first excited
band, has also mixing from the ground-state and other
configurations.  Although all the states are mixed, but it is
possible to identify them at low-spin values as they have one
predominant component. For high-spin states, the bands are 
highly mixed, and it is difficult to identify them. 


The complete band structures for the lowest and the first excited state, obtained after
diagonalization of the shell model Hamiltonian, are depicted
in Figs.~\ref{expe1}, \ref{expe11}, and \ref{expe2} for Pr- isotopes and
Figs.~\ref{expe101} and \ref{expe22}  for Pm- isotopes and
are compared with the experimental data, wherever available. For most
of the nuclei, the ground-state band, except for $^{127}$Pm and
$^{129}$Pm, are known up to quite high-spin and the TPSM energies
are noted to be in good agreement with these known level energies. 
 Some preliminary TPSM results for $^{135}$Pm were presented in the experimental work  \cite{fs21} and it was shown that the results agreed remarkably well with the data.
The energy values for all the states have been specified in the Figs.~\ref{expe1} - \ref{expe22} ,
which shall be useful to make comparisons with future experimental measurements,
and as well as with other theoretical studies. 

We now turn our discussion to the band crossing features observed 
in the studied Pr- and Pm-isotopes. As already stated in the
introduction, the observed band crossing features in some of these
isotopes could not be explained using the standard CSM approach 
and it was necessary to employ the self-consistent TRS approach
to shed light on the anomalous band crossing features.
Further, in the earlier PSM study of the odd-proton isotopes, only neutron
aligned states were considered in the basis space \cite{ysm}. However, it was
evident from the earlier analysis \cite{cm98} that neutron and proton alignments compete,
and it is imperative to include both two-neutron and two-proton
aligned configurations in the basis space. In the present work, both these
configurations have been included and in the following, we shall
present the results of alignments and moments of inertia.
Alignment, $i_x$ and the dynamic moment of
inertia, $J^{(2)}$ have been evaluated using the standard
expressions \cite{RB97}.
These quantities are displayed in Figs.~\ref{ali1} and \ref{ali2} for the
Pr-isotopes. For $^{125}$Pr, $^{127}$Pr and $^{129}$Pr, both expt. and TPSM
deduced $i_x$ depict an increasing trend with spin and band crossing
is not evident from this plot. As we shall see below that $J^{(2)}$,
which is more sensitive to changes in the alignment, depicts band
crossing features. $i_x$ plots for $^{131}$Pr, $^{133}$Pr, $^{135}$Pr
and $^{137}$Pr show 
upbends, which is indicative of a band crossing having large
interaction strength between the ground-state and the aligned band. 
It is evident from the figure that TPSM results agree fairly well with those deduced from the experimental data. 

For $^{125-133}$Pr isotopes, $J^{(2)}$ in Fig.~\ref{ali2}
depict upbends between spin values of I $=31/2$, and $35/2$. The upbend is
a clear indication of the change in the configuration along the yrast
band and is a signature of the band
crossing phenomenon. For $^{135}$Pr, the upbend in $J^{(2)}$ is noticed
at a higher angular momenta of I $=41/2$ and in the case of $^{137}$Pr,
the discontinuity in $J^{(2)}$ is observed at a lower
angular-momentum. The $J^{(2)}$ evaluated from the measured energies 
depicts a larger enhancement as  compared to the TPSM predicted value.

For the isotopes of Pm from A=127 to 135, $i_x$ plotted in
Fig.~\ref{ali3} depicts an increasing
trend with spin. For A=127 and 129, experimental quantities are not 
available, but for A=131, 133 and 135, TPSM values are in good
agreement with the data. For $^{137}$Pm and $^{137}$Pm, two upbends
are predicted by TPSM calculations, and for $^{139}$Pm, both the
upbends are observed in the experimental data. $J^{(2)}$ calculated for the
studied Pm-isotopes are depicted in Fig.~\ref{ali4} and are noted to be
in good agreement with the known experimental quantities, except that
for the isotopes of $^{135}$Pm and $^{137}$Pm, the TPSM calculated upbends are
smoother in comparison to the experimental quantities. 

It has been demonstrated using the self-consistent TRS model \cite{cm98}
that alignments for the studied odd-proton Pr- and
Pm-isotopes are quite complicated with considerable mixing between neutron and
proton configurations. In the band diagrams of Figs.~\ref{bd3} and \ref{bd4}, the
alignment is  either due to protons or neutrons as the energies are plotted
before configuration mixing. In order to investigate the mixing
between the neutron and proton configurations, the wavefunction
amplitudes are depicted in Fig.~\ref{wf3} for $^{127}$Pr and $^{131}$Pm,
which were studied in detail in Ref.~\cite{cm98}. It is quite evident
from the figure that band crossing is not entirely due to the
alignment of two neutrons, but also has a significant contribution 
from the aligned proton configuration. For heavier isotopes, the
situation is reversed with proton contribution larger than the neutron
one.  Therefore, present work substantiates the TRS prediction that
alignments for odd-proton Pr- and Pm-isotopes are quite complicated
with mixing between the proton and neutron aligned configurations. This
is primarily due to the reason that both aligned protons and neutrons occupy the
same intruder orbital, $1h_{11/2}$. 

\section{Summary and conclusions}

In the present work, the TPSM approach for odd-proton nuclei has been
generalized to include three-proton and three-proton coupled to two-neutron
quasiparticle configurations. This extension allows the application of
the TPSM approach to high-spin band structures observed in odd-proton
system. In the earlier version, only one-proton and one-proton coupled to
two-neutron configurations were considered and this limited the
application of the TPSM approach. In some odd-proton Pr- and Pm-
isotopes, anomalous band crossing features were reported using the
standard CSM analysis. It was demonstrated, using a more realistic TRS
approach  in which pairing and deformation properties were obtained
self-consistently, that first band crossing in some Pr- and Pm-
isotopes also contained a large contribution from the proton
configuration. Normally, it is expected that in odd-proton system, the 
proton crossing is blocked for the yrast band and the first crossing is due to the
alignment of two-neutrons.  It has been shown using the extended basis space that for lighter Pr-
and Pm-isotopes, the band crossings is dominated by the alignment of 
neutrons. However, for heavier isotopes, it has been shown  that
first band crossing has dominant contribution from the aligned
protons. The present work also confirmed the TRS prediction that
band crossings in Pr- and Pm-isotopes have mixed neutron and proton
aligned configurations. 

Further, we have  explored the possibility of observing $\gamma$-bands in the
studied odd-mass systems. $\gamma$-bands are quite scarce in odd-mass systems
and have been observed only in a few nuclei. In comparison,
the $\gamma$-bands in even-even systems have been observed, not only based
on the ground-state, but have also been identified built on two-quasiparticle excited
configurations. In even-even Ce- and Nd-isotopes,
several s-bands are observed and it was shown that some of these
s-bands are as a matter of fact, $\gamma$-bands built on the
two-quasiparticle states. 
Since Ce- and Nd-isotopes are even-even cores of Pr- and Pm-isotopes,
it is expected that they should also depict some features of the
even-even cores. It has been shown in the present work that
heavier Pr- and Pm-isotopes are the best candidates to observe the $\gamma$-bands based
on three-quasiparticle configurations. We have provided the excitation
energies of the band heads of these structures which shall be helpful in identifying
them in future experimental studies.

\section{ACKNOWLEDGEMENTS}
The authors  would like to acknowledge Department of Science and
Technology  ( Govt. of India ) for providing financial assistance under the 
Project No.$CRG/2019/004960$ to carry out a part of the present research work.

\end{document}